\documentclass[12pt]{article}
\usepackage{euscript,amsmath,amssymb,amsfonts}

\newcounter{appen}

\newcommand\Sum{\sideset{}{^{\,\lower1mm\hbox{$\oplus$}}}\sum}
\input{tcilatex}

\begin{document}

\title{{\Large Dirac Hamiltonian with superstrong Coulomb field}}
\author{B.L. Voronov\thanks{%
Lebedev Physical Institute, Moscow, Russia; e-mail: voronov@lpi.ru}, D.M.
Gitman\thanks{%
Institute of Physics, University of Sao Paulo, Brazil; e-mail:
gitman@dfn.if.usp.br}, and I.V. Tyutin\thanks{%
Lebedev Physical Institute, Moscow, Russia; e-mail: tyutin@lpi.ru}}
\maketitle

\begin{abstract}
We consider the quantum-mechanical problem of a relativistic Dirac particle
moving in the Coulomb field of a point charge $Ze$. In the literature, it is
often declared that a quantum-mechanical description of such a system does
not exist for charge values exceeding the so-called critical charge with $%
Z=\alpha ^{-1}=137$ based on the fact that the standard expression for the
lower bound state energy yields complex values at overcritical charges. We
show that from the mathematical standpoint, there is no problem in defining
a self-adjoint Hamiltonian for any value of charge. What is more, the
transition through the critical charge does not lead to any qualitative
changes in the mathematical description of the system. A specific feature of
overcritical charges is a non uniqueness of the self-adjoint Hamiltonian,
but this non uniqueness is also characteristic for charge values less than
the critical one (and larger than the subcritical charge with $Z=(\sqrt{3}%
/2)\alpha ^{-1}=118$). We present the spectra and (generalized)
eigenfunctions for all self-adjoint Hamiltonians. The methods used are the
methods of the theory of self-adjoint extensions of symmetric operators and
the Krein method of guiding functionals. The relation of the constructed
one-particle quantum mechanics to the real physics of electrons in
superstrong Coulomb fields where multiparticle effects may be of crucial
importance is an open question.
\end{abstract}


\section{Introduction}

It is common knowledge that the complete sets of solutions of relativistic
wave equations (like Klein-Gordon equation, Dirac equation, and so on) when
used in quantizing the corresponding (scalar, spinor, and so on) free fields
allow interpreting the corresponding quantum theories in terms of particles
and antiparticles~\cite{1}. The space of quantum states of each such a field
is decomposed into sectors with a definite number of particles (the vacuum,
one-particle sector, and so on), each sector is stable under time evolution.
A description of the one-particle sector of a free QFT can be formulated as
a relativistic quantum mechanics where the corresponding relativistic wave
equations play the role of the Schr\"{o}dinger equation and their solutions
are interpreted as wave functions of particles and antiparticles. In QED
(and some other models), the concept of the external electromagnetic field
is widely and fruitfully used. It can be considered an approximation in
which a ``very intensive'' part of the electromagnetic field is treated
classically and is not subjected to any back reaction of the rest of the
system.The Dirac equation with such a field plays an important role in QED
with an external field. Of special interest are the cases where an external
field allows exactly solving the Dirac equation. There are a few of such
exactly solvable cases of physically interesting external electromagnetic
fields, see, e.g.,~\cite{2}. They can be classified into groups such that
the Dirac equations with fields of each group have a similar interpretation.

The constant uniform magnetic field, the plane-wave field, and their
parallel combination form a first group, the fields of this group do not
violate the vacuum stability (do not create particles from the vacuum). The
exact solutions of the Dirac equation with such fields form complete systems
and can be used in the quantization procedure providing a particle
interpretation for a quantum spinor field in the corresponding external
background. This allows constructing an approximation where the interaction
with the external field is taken into account exactly, while the interaction
with the quantized electromagnetic field is treated perturbatively. Such an
approach to QED with external fields of the first group is known as the
Furry picture, see, e.g.,~\cite{3,1}. In the Furry picture, the state space
of the quantum theory of the spinor field with the external fields is
decomposed into sectors with a definite number of particles, each sector is
stable under time evolution similarly to the zero external-field case. The
description of the one-particle sector also can be formulated as a
relativistic quantum mechanics~\cite{4}. We note that the solutions of the
Dirac equation with an uniform magnetic field provide a basis for the
quantum synchrotron radiation theory~\cite{5}, and the solutions of the
Dirac equation with the plane-wave field are widely used for calculating the
quantum effects when electrons and other particles of spin one-half move in
laser beams~\cite{6}.

A uniform electric field and some other electromagnetic fields violate the
vacuum stability. A literal application of the above approach to
constructing the Furry picture in QED with such fields is fails. However, it
was demonstrated that existing exact solutions of the Dirac equation with
electric-type fields form complete sets and can be used for describing a
vsriety of quantum effects in such fields, in particular, the
electron-positron pair production from the vacuum~\cite{7}. Moreover, these
sets of solutions form a basis for constructing a generalized Furry picture
in QED with external fields violating the vacuum stability, see~\cite{8}. It
should be noted that the one-particle sector in such external fields is
unstable under time evolution, and therefore, the corresponding quantum
mechanics of a spinning particle cannot be constructed in principle.

A study of the Dirac equation with a singular external Aharonov-Bohm field,
and with some additional fields revealed problems of a new type. Although
some sets of exact solutions of the Dirac equation with such fields can be
found, the problem of the completeness of these sets arises. This problem is
related to the fact that the Dirac Hamiltonian with the singular external
Aharonov-Bohm field should be additionally specified for it can be treated
as a self-adjoint (s.a. in what follows) quantum-mechanical operator. It can
be shown (for a review, see~\cite{9}) that in this case, there exists a
family of s.a. Hamiltonians which are constructed by methods of the theory
of s.a. extensions of symmetric operators dating back to von Neumann~\cite%
{10}. Each s.a. Hamiltonian yields a complete set of solutions which can be
used for constructing the Furry picture in QED with the singular external
Aharonov-Bohm field (this field does not violate the vacuum stability).

The Dirac equation with the Coulomb field, and with some additional fields,
has always been of particular interest. The Coulomb field is even referred
to as a ``microscopic external field'' to underline its qualitative
distinction from the above-mentioned external fields which are sometimes
referred to as ``macroscopic'' ones. Until recently, the commonly accepted
view on the situation in the theory was the following. The Dirac equation
for an electron of charge\footnote{$e=4,803\times 10^{-10}CGSE$ is the
magnitude of the electron charge.} $-e$ in an external Coulomb field created
by a positive point-like electric charge $Ze$ of a nucleus of atomic number%
\footnote{$\alpha =e^{2}/\hbar c$ is the fine structure constant.} $Z\leq
1/\alpha =137$ is solved exactly, has a complete set of solutions, and
allows constructing a relativistic theory of atomic spectra which is in
agreement with experiment~\cite{11}. This field does not violate the vacuum
stability, therefore, the Furry picture can be constructed, and there exists
the relativistic quantum mechanics of the spinning particle in such a
Coulomb field. As for the Dirac equation with the Coulomb field with $Z>137$%
, it was considered inconsistent and physically meaningless~\cite{12}--\cite%
{15}. One of the standard arguments is that the formula for the lower $%
1S_{1/2}$ energy level, 
\begin{equation*}
E_{1s}=mc^{2}\sqrt{1-\left( Z\alpha \right) ^{2}}\,,
\end{equation*}%
formally gives imaginary eigenvalues for the Dirac Hamiltonian with $Z>137$.
On the one hand, the question of the consistency of the Dirac equation with
the Coulomb field with $Z>137$ has a pure theoretical (mathematical)
interest, on the other hand, it is concerned with the question of the
electron structure of atoms of atomic numbers $Z>1/\alpha $ and especially
of atoms with nuclei of supercritical charge $Ze>170e$ . The latter question
is of fundamental importance. The formulation of QED cannot be considered
really completed until an exhaustive answer to this question is given.
Although nuclei of so large electric charges can hardly be synthesized%
\footnote{%
At present, the maximum $Z=118.$}, the existing heavy nuclei can imitate the
supercritical Coulomb fields at collision. Nuclear forces can hold the
colliding nuclei together for $10^{-19}s$ or more. This time is enough to
reproduce effectively the experimental situation where the electron
experiences the supercritical Coulomb field~\cite{15}. Several groups of
researchers attacked the problem of the behavior of the electron in the
supercritical Coulomb field, see~\cite{15},~\cite{16}. The difficulty of the
imaginary spectrum in the case of $Z>137$ was attributed to an inadmissible
singularity of the supercritical Coulomb field for a relativistic electron 
\footnote{%
An equation for the radial components of wave functions has the form of the
nonrelativistic Schr\"{o}dinger equation with an effective potential with
the $r^{-2}$ singularity at the origin, which is associated with ``a fall to
the center''.}. It was believed that this difficulty can be eliminated if a
nucleus of some finite radius $R$ is considered. It was shown that with
cutting off the Coulomb potential with $Z<170$ at a radius $R\sim 1,2\times
10^{-12}cm$, the Dirac equation has physically meaningful solutions \cite{17}%
. But even in the presence of the cut off, another difficulty arises at $%
Z\sim 170$. Namely, the lower bound state energy descends to the upper
boundary $E=-mc^{2}$ of the lower continuum, and it is generally agreed that
in such a situation, the problem can no longer be considered a one-particle
one because of the electron-positron pair production, which in particular
results in a screening of the Coulomb potential of the nucleus.
Probabilities of the particle production in the heavy-ion collisions were
calculated in the framework of this conception~\cite{15}. Unfortunately,
experimental conditions for verifying the corresponding predictions are
unavailable at present.

In this paper, we turn back to the problem of the consistency of the Dirac
equation with the Coulomb field with no cut off and with arbitrary nucleus
charge values( with arbitrary $Z$). Our point of view is that the
above-mentioned difficulties with the spectrum for $Z>137$ do not arise if
the the Dirac Hamiltonian is correctly defined as a s.a. operator. We
present a rigorous treatment of all the aspects of this problem including a
complete spectral analysis of the model based on the theory of s.a.
extensions of symmetric operators and the Krein method of guiding
functionals. We show that from the mathematical standpoint, the definition
of the Dirac Hamiltonian as a s.a. operator for arbitrary $Z$ presents no
problem. What's more, the transition from the noncritical charge region to
the critical one does not lead to qualitative changes in the mathematical
description of the system. A specific feature of the overcritical charges is
a non uniqueness of the s.a. Dirac Hamiltonian, but this non uniqueness is
characteristic even for charge values less than the critical one (and larger
than the subcritical value with $Z=(\sqrt{3}/2)\alpha ^{-1}=118$).
Presenting a rigorous treatment of the problem, we also compare it with the
conventional physical approach to constructing a quantum-mechanical
description of the relativistic Coulomb system. It turns out that many of
mathematical results can be obtained from physical considerations except the
important property of completeness for the eigenfunctions of the
Hamiltonian. The obtained complete sets of solutions can be used to
construct the Furry picture in QED. However, it is unclear whether the
neglect of the radiative interaction in such a Furry picture is a good zero
approximation to describing quantum effects in QED with the Coulomb field
with no cut off and with arbitrary nucleus charge values. In other words, a
relevance of the constructed quantum-mechanics with an energy spectrum
unbounded from below to the real physics of an electron in supercritical
Coulomb fields where multiparticle effects may be of crucial importance is
an open question.

The paper is organized as follows. In Sec.~\ref{sec2}, we present basic
facts and formulas clarifying the formulation of the problem and reduce the
problem of constructing a s.a. Dirac Hamiltonian with an external Coulomb
field in the whole Hilbert space to the problem of constructing s.a.
one-dimensional radial Hamiltonians. In Sec.~\ref{sec3}, we cite expressions
for the general solution of the radial equations and some particular
solutions of these equations used in the following. In Subsec.~\ref{sec4.1}
and Appendix~\ref{appA}, we outline procedures for constructing s.a.
extensions of symmetric differential operators and their spectral analysis.
In Subsecs.~\ref{sec4.2}--\ref{sec4.5}, we construct s.a. Dirac Hamiltonians
in all four possible charge regions and find their spectra and the
corresponding complete sets of eigenfunctions.

\section{Setting the problem}

\label{sec2}

We consider the Dirac equation for a particle of charge $q_{1}$ moving in
the external Coulomb field of a point-like charge $q_{2}$; for an electron
in a hydrogen-like atom, we have $q_{1}=-e\,,$ $q_{2}=Ze.$ We choose the
electromagnetic potentials for such a field in the form 
\begin{equation*}
A_{0}=\frac{q_{2}}{r}\,,\;A_{k}=0\,.
\end{equation*}%
The Dirac equation with this field, being written in the form of the Schr%
\"{o}dinger equation, is\footnote{%
We use the bold letters for there-vectors and the standard representation
for $\gamma$-matrices where 
\begin{equation*}
\mathbf{\alpha }=\left( 
\begin{array}{lr}
0 & \mathbf{\sigma } \\ 
\mathbf{\sigma } & 0%
\end{array}%
\right) \,,\;\beta =\gamma ^{0}=\left( 
\begin{array}{lr}
I & 0 \\ 
0 & -I%
\end{array}%
\right) \,,\;\mathbf{\Sigma }=\left( 
\begin{array}{cc}
\mathbf{\sigma } & 0 \\ 
0 & \mathbf{\sigma }%
\end{array}%
\right) \,,
\end{equation*}%
and $\mathbf{\sigma }=\left( \sigma ^{1},\sigma ^{2},\sigma ^{3}\right) $
are the Pauli matrices. We use the notation $\mathbf{\sigma p=}\sigma
^{k}p^{k},\;\mathbf{\sigma r=}\sigma ^{k}x^{k},\;$ and so on. We set $\hbar
=c=1$ in what follows.}: 
\begin{equation*}
i\frac{\partial \Psi \left( x\right) }{\partial t}=\check{H}\Psi \left(
x\right) \,,\;x=\left( x^{0},x^{k}\right) =\left( t,\mathbf{r}\right) \,,
\end{equation*}%
where $\Psi \left( x\right) =\{\psi _{\alpha }(x)\}$ is a bispinor and the
Dirac Hamiltonian$\;\check{H}$\ is given by 
\begin{equation}
\check{H}=\mathbf{\alpha \check{p}}+m\beta -\frac{q}{r}=\left( 
\begin{array}{cc}
m-q/r & \mathbf{\sigma \check{p}} \\ 
\mathbf{\sigma \check{p}} & -m-q/r%
\end{array}%
\right) ,  \label{1.3}
\end{equation}%
$m$ is the fermion mass, $\mathbf{\check{p}}=\left( \check{p}^{k}=-i\partial
_{k}\right) ,$ and $q=-q_{1}q_{2}$; for an electron in a hydrogen-like atom,
we have $q=Z\alpha $. For brevity, we call the coupling constant $q$ the
charge. We restrict ourselves to the case of $q>0$, because the results for
the case of $q<0$ can be obtained by the charge conjugation transformation.

At this initial stage of setting the problem, the Hamiltonian $\check{H}$
and other operators are considered as formally s.a. differential operators,
or s.a. differential expressions\footnote{%
S.a. by Lagrange in the mathematical terminology, or formally s.a. in the
physical terminology.}, as we will say~\cite{18}, which is denoted by the
turned hat $\vee $ above the corresponding letter. They become
quantum-mechanical operators after a specification of their domains in the
Hilbert space $\mathcal{H}$ of bispinors $\Psi(\mathbf{r})$, 
\begin{equation*}
\mathcal{H}=\sideset{}{^{\,\lower1mm\hbox{$\oplus$}}}\sum_{\alpha=1}^{4}%
\mathcal{H}_{\alpha }, \qquad \mathcal{H}_{\alpha}=L^{2}(R^{3})
\end{equation*}
then the symbol $\vee $ is replaced by the conventional symbol $\wedge $
over the same letter. In what follows, we distinguish differential
expressions $\check{f}$ and operators $\hat{f}$ and call $\hat{f}$ the
operator associated with the differential expression $\check{f}$.

The purposes of this paper are constructing a s.a. Hamiltonian $\hat{H}$
associated with $\check{H}$, which primarily means indicating a domain of $%
\hat{H}$, and then finding its spectrum and eigenfunctions.

We note that in the physical literature, the eigenvalue problem is
conventionally considered directly in terms of the s.a. differential
expression $\check{H}\;$as the eigenvalue problem for the differential
equation $\check{H}\Psi _{E}\left( \mathbf{r}\right) =E\Psi _{E}\left( 
\mathbf{r}\right) $, the stationary Dirac equation, without any reference to
the domain of the Hamiltonian\footnote{%
In fact, some natural domain is implicitly implied.}.\ It is solved by
separating variables based on the rotation symmetry of the problem. The
rotation symmetry is conventionally treated in terms of s.a. differential
expressions as follows.

The Dirac Hamiltonian $\check{H}$ formally commutes with the total angular
momentum $\mathbf{\check{J}}=\mathbf{\check{L}}+\frac{1}{2}\mathbf{\Sigma }$%
,\ where $\mathbf{\check{L}}=\left[ \mathbf{r}\times \mathbf{\check{p}}%
\right] $\ is the orbital angular momentum operator and $\frac{1}{2}\mathbf{%
\Sigma }$\textbf{\ }is the spin angular momentum operator, and the so-called
spin operator $\check{K}$, 
\begin{equation*}
\check{K}=\beta \left[ 1+\left( \mathbf{\Sigma \check{L}}\right) \right]
=\left( 
\begin{array}{cc}
\check{\varkappa} & 0 \\ 
0 & -\check{\varkappa}%
\end{array}%
\right) ,\;\check{\varkappa}=1+\left( \mathbf{\sigma \check{L}}\right) .
\end{equation*}%
The differential expressions $\check{H}$, $\mathbf{\check{J}}^{2}$, $\check{J%
}_{3}$, and $\check{K}$ are considered a complete set of commuting
operators, which allows separating the angular and radial variables and
reducing the total stationary Dirac equation to the radial stationary Dirac
equation with a fixed angular momentum, its $z$-axis projection, and a spin
operator eigenvalue.

We here present a treatment of the problem which is proper from the
functional analysis standpoint. We construct a s.a. Hamiltonian $\hat{H}$
based on the theory of s.a. extensions of symmetric operators and on the
rotation symmetry of the problem. This means that we first define a
rotationally invariant symmetric operator $\hat{H}^{(0)}$ associated with
s.a. differential expression $\check{H}\;$(\ref{1.3}), which is rather
simple, and then find its rotationally invariant s.a. extensions. Because
the coefficient functions of $\check{H}$ are smooth out of the origin, we
choose the space of smooth bispinors with a compact support \footnote{%
We thus avoid troubles associated with a behavior of wave functions at
infinity.} for the domain $D_{H^{(0)}}$ of $\hat{H}^{(0)}$. To avoid
troubles with the $1/r$ singularity of the potential at the origin, we
additionally require that all bispinors in $D_{H^{(0)}}$ vanish near the
origin\footnote{%
Strictly speaking, we thus leave room for $\delta $-like terms in the
potential.}. The operator $\hat{H}^{(0)}$is thus defined by 
\begin{equation*}
\hat{H}^{(0)}:\left\{ 
\begin{array}{l}
D_{H^{(0)}}=\{\Psi (\mathbf{r}):\;\psi _{\alpha }(\mathbf{r})\in
D(R^{3});\;\psi _{\alpha }(\mathbf{r})=0,\;\mathbf{r\in }U_{\varepsilon }\},
\\ 
\hat{H}^{(0)}\Psi (\mathbf{r})=\check{H}\Psi (\mathbf{r}),%
\end{array}%
\right.
\end{equation*}%
where $D(R^{3})$ is the space of smooth functions in $R^{3}$ with a compact
support and $U_{\varepsilon }$ is some vicinity of the origin which is
generally different for different bispinors. The domain $D_{H^{(0)}}$ is
dense in $\mathcal{H}$, $\overline{D_{H^{(0)}}}=\mathcal{H}$, and the
symmetricity of $\hat{H}^{(0)}$ is easily verified by integrating by parts.

We now take the rotational invariance into account. The operator $\hat{H}%
^{(0)}$ evidently commutes with the s.a. angular momentum operator $\mathbf{%
\hat{J}}=\{\hat{J}_{k}\}$ and the s.a. operator $\hat{K}$ associated with
the respective differential expressions $\mathbf{\check{J}}$ and $\check{K}$%
. The operators $\hat{J}_{k}$ are defined as generators of the unitary
representation of the rotation group $\mathrm{Spin}(3)\mathcal{\ }$in the
Hilbert space $\mathcal{H}$. The Hilbert space $\mathcal{H}$ is represented
as a direct orthogonal sum, 
\begin{equation}
\mathcal{H}=\sideset{}{^{\,\lower1mm\hbox{$\oplus$}}}\sum_{j,\zeta}\mathcal{H%
}_{j,\zeta },\;j=1/2,3/2,...,\;\zeta =\pm 1,  \label{R.1}
\end{equation}%
of subspaces $\mathcal{H}_{j,\zeta }$. The subspaces $\mathcal{H}_{j,\zeta }$
reduce the operators $\mathbf{\hat{J}}^{2},\;J_{k},$ and $\hat{K}$,\footnote{%
This means that the operators $\mathbf{\hat{J}}^{2}$, $\hat{J}_{3}$, and $%
\hat{K}$ commute with the projectors to the subspaces $\mathcal{H}_{j,\zeta}$%
, see~\cite{18}.} 
\begin{equation*}
\mathbf{\hat{J}}^{2}=\sideset{}{^{\,\lower1mm\hbox{$\oplus$}}}\sum_{j,\zeta }%
\mathbf{\hat{J}}_{j,\zeta }^{2},\;\hat{J}_{k}=\sideset{}{^{\,\lower1mm%
\hbox{$\oplus$}}}\sum_{j,\zeta }\hat{J}_{kj,\zeta },\;\hat{K}=%
\sideset{}{^{\,\lower1mm\hbox{$\oplus$}}}\sum_{j,\zeta }\hat{K}_{j,\zeta }.
\end{equation*}%
In its turn, the subspaces $\mathcal{H}_{j,\zeta }$ are finite direct sums
of the subspaces 
\begin{equation}
\mathcal{H}_{j,\zeta }=\sideset{}{^{\,\lower1mm\hbox{$\oplus$}}}\sum_{M}%
\mathcal{H}_{j,M,\zeta },\;M=-j,-j+1,...,j.  \label{R.1ab}
\end{equation}%
The subspace $\mathcal{H}_{j,M,\zeta }$ is the subspace of bispinors $\Psi
_{j,M,\zeta }\left( \mathbf{r}\right) $ of the form 
\begin{equation}
\Psi _{j,M,\zeta }\left( \mathbf{r}\right) =\frac{1}{r}\left( 
\begin{array}{c}
\Omega _{j,M,\zeta }(\theta ,\varphi )f\left( r\right) \\ 
i\Omega _{j,M,-\zeta }(\theta ,\varphi )g\left( r\right)%
\end{array}%
\right) \,,  \label{R.1a}
\end{equation}%
where $\Omega _{j,M,\zeta }(\theta ,\varphi )$ are spherical spinors and $%
f\left( r\right) $ and $g\left( r\right) $ are radial functions (the factors 
$1/r$ and $i$ are introduced for convenience). The subspaces $\mathcal{H}%
_{j,M,\zeta }$ are the eigenspaces of the operators $\mathbf{\hat{J}}^{2}$, $%
\hat{J}_{3}$, and $\hat{K}$: 
\begin{align*}
& \mathbf{\hat{J}}^{2}\Psi _{j,M,\zeta }\left( \mathbf{r}\right) =j(j+1)\Psi
_{j,M,\zeta }\left( \mathbf{r}\right) ,\;\hat{J}_{3}\Psi _{j,M,\zeta }\left( 
\mathbf{r}\right) =M\Psi _{j,M,\zeta }, \\
& \hat{K}\Psi _{j,M,\zeta }\left( \mathbf{r}\right) =-\zeta (j+1/2)\Psi
_{j,M,\zeta }\left( \mathbf{r}\right) ,
\end{align*}%
and evidently reduce the operators $\mathbf{\hat{J}}^{2}$, $\hat{J}_{3}$,
and $\hat{K}$. In the physical language, decomposition (\ref{R.1}), (\ref%
{R.1ab}) corresponds to the expansion of bispinors $\Psi \left( \mathbf{r}%
\right) \in \mathcal{H}$ in terms of the eigenfunctions of the commuting
operators $\mathbf{\hat{J}}^{2}$, $\hat{J}_{3}$, and $\hat{K}$, which allows
separating variables in the equations for eigenfunctions.

We note that the reductions $\hat{J}_{1,2j,\zeta}$ of the operators $\hat{J}%
_{1,2}$ to the subspaces $\mathcal{H}_{j,\zeta}$ are bounded operators.

The following fact is basic for us. Let $\mathcal{L}^{2}(0,\infty)$ be the
Hilbert space of doublets $F(r)$, 
\begin{equation*}
F(r)=\left( 
\begin{array}{c}
f(r) \\ 
g(r)%
\end{array}
\right) ,
\end{equation*}
with the scalar product 
\begin{equation*}
\left( F_{1},F_{2}\right) =\int_{0}^{\infty}drF_{1}^{+}\left( r\right)
F_{2}\left( r\right) =\int_{0}^{\infty}dr\left[ \overline{f_{1}\left(
r\right) }f_{2}\left( r\right) +\overline{g_{1}\left( r\right) }g_{2}\left(
r\right) \right] ,
\end{equation*}
such that $\mathcal{L}^{2}(0,\infty)=L^{2}(0,\infty){}^\oplus
L^{2}(0,\infty) $. Then formula (\ref{R.1a}) and the relation 
\begin{equation*}
||\Psi_{j,M,\zeta}||^{2}=\int d\mathbf{r}\Psi_{j,M,\zeta}^{+}(\mathbf{r}%
)\Psi_{j,M,\zeta}(\mathbf{r})=\int dr[|f(r)|^{2}+|g(r)|^{2}]
\end{equation*}
show that the Hilbert space $\mathcal{H}_{j,M,\zeta}$ is unitary equivalent
to the Hilbert space $\mathcal{L}^{2}(0,\infty)$: 
\begin{equation*}
F=U_{j,M,\zeta}\Psi_{j,M,\zeta},\;\Psi_{j,M,\zeta}=U_{j,M,\zeta}^{-1}F,
\end{equation*}
the explicit form of the unitary operator $U$ is defined by (\ref{R.1a}).

The rotational invariance of $\hat{H}^{(0)}$ is equivalent to the following
statement.

1) The subspaces $\mathcal{H}_{j,M,\zeta}$ reduce this operator, such that $%
\hat{H}^{(0)}$ is represented as a direct orthogonal sum of its parts $%
\hat{H}_{j,\zeta}^{(0)}$ and $\hat{H}_{j,M,\zeta}^{(0)}$ that are the
reductions of $\hat{H}^{(0)}$ to the respective $\mathcal{H}_{j,\zeta}$ and $%
\mathcal{H}_{j,M,\zeta}$, 
\begin{equation}
\hat{H}^{\left( 0\right) }\mathbf{=}\sideset{}{^{\,\lower1mm\hbox{$\oplus$}}}%
\sum_{j,\zeta}\hat{H}_{j,\zeta }^{\left( 0\right) },\;\hat{H}%
_{j,\zeta}^{\left( 0\right) }\mathbf{=}\sideset{}{^{\,\lower1mm\hbox{$%
\oplus$}}}\sum_{M}\hat{H}_{j,M,\zeta}^{\left( 0\right) }.  \label{R.5}
\end{equation}

Each part $\hat{H}_{j,M,\zeta }^{\left( 0\right) }$ is a symmetric operator
in the Hilbert space $\mathcal{H}_{j,M,\zeta }$. Each symmetric operator $%
\hat{H}_{j,M,\zeta }^{\left( 0\right) }$ in the subspace $\mathcal{H}%
_{j,M,\zeta }$ evidently induces a symmetric operator $\hat{h}_{j,\zeta
}^{\left( 0\right) }$ in the Hilbert space $\mathcal{L}^{2}(0,\infty )$, 
\begin{equation*}
\hat{h}_{j,\zeta }^{\left( 0\right) }F=U_{j,M,\zeta }\hat{H}_{j,M,\zeta
}^{\left( 0\right) }\Psi _{j,M,\zeta },
\end{equation*}%
such that $\hat{h}_{j,\zeta }^{\left( 0\right) }=U_{j,M,\zeta }\hat{H}%
_{j,M,\zeta }^{\left( 0\right) }U_{j,M,\zeta }^{-1}$, and $\hat{h}_{j,\zeta
}^{\left( 0\right) }$ is given by 
\begin{equation}
\hat{h}_{j,\zeta }^{\left( 0\right) }:\left\{ 
\begin{array}{l}
D_{h_{j,\zeta }^{\left( 0\right) }}=\mathcal{D}\left( 0,\infty \right) , \\ 
\hat{h}_{j,\zeta }^{\left( 0\right) }F(r)=\check{h}_{j,\zeta }F(r),%
\end{array}%
\right.  \label{R.3}
\end{equation}%
where $\mathcal{D}(0,\infty )=D(0,\infty ){}^\oplus D(0,\infty )$, $%
D(0,\infty ) $ is the standard space of smooth functions on $(0,\infty )$
with a compact support, 
\begin{equation*}
D\left( 0,\infty \right) =\left\{ f\left( r\right) :f\left( r\right) \in
C^{\infty },\;\mathrm{supp}f\subset \lbrack a,b],\;0<a<b<\infty \right\} ,
\end{equation*}%
the segment $[a,b]$ is generally different for different $f$, and the s.a.
radial differential expression $\check{h}_{j,\zeta }$ is given by 
\begin{equation}
\check{h}_{j,\zeta }=-i\sigma ^{2}\frac{d}{dr}+\frac{\varkappa }{r}\sigma
^{1}-\frac{q}{r}+m\sigma ^{3},  \label{R.4}
\end{equation}%
where $\varkappa =\zeta (j+1/2)$.\footnote{%
We note that $\mathcal{D}\left( 0,\infty \right) $ is dense\ in $\mathcal{L}%
^{2}(0,\infty )$, because, as is known, $D\left( 0,\infty \right) $ is dense
in $L^{2}(0,\infty )$, and the symmetricity of $\hat{h}_{jM\zeta }^{\left(
0\right) }$ is easily verified by integrating by parts, which confirms the
above assertion.}

2) The differential expression $\check{h}_{j,\zeta }$\ and consequently,
with taking (\ref{R.3}) into account, the operator $\hat{h}_{j,\zeta
}^{\left( 0\right) }$ with fixed $j$ and $\zeta $ is independent of $M$.
This fact is equivalent to the commutativity of the operator $\hat{H}%
^{\left( 0\right) }$ with the operators $\hat{J}_{1}$ and $\hat{J}_{2}$, or
more precisely, to the commutativity of the operator $\hat{H}_{j,\zeta
}^{\left( 0\right) }$ with the operators $\hat{J}_{1,2j,\zeta }$. In the
physical terminology, $\check{h}_{j,\zeta }$ is called the radial
Hamiltonian, but strictly speaking, the radial Hamiltonian is a s.a.
operator $\hat{h}_{j,\zeta }$ associated with $\check{h}_{j,\zeta }$.

In what follows, by the rotational invariance of any operator $\hat{f}$ , we
mean the fulfilment of the following conditions:

1) the reducibility of this operator by the subspaces $\mathcal{H}%
_{j,M,\zeta }$ and therefore, by $\mathcal{H}_{j,\zeta }$, such that formula
similar to (\ref{R.5}) holds for the operator $\hat{f}$ ;

2) the commutativity of its parts $\hat{f}_{j,\zeta }$ with the bounded
operators $\hat{J}_{1,2j,\zeta }$ for fixed $j,\zeta $.

Let $\hat{h}_{j,\zeta }$ be a s.a. extension of the symmetric operator $%
\hat{h}_{j,\zeta }^{\left( 0\right) }$\ in $\mathcal{L}^{2}(0,\infty )$. It
evidently induces s.a. extensions $\hat{H}_{j,M,\zeta }$ of the symmetric
operators $\hat{H}_{j,M,\zeta }^{\left( 0\right) }$ in the subspaces $%
\mathcal{H}_{j,M,\zeta }$, 
\begin{equation}
\hat{H}_{j,M,\zeta }=U_{j,M,\zeta }^{-1}\hat{h}_{j,\zeta }U_{j,M,\zeta },
\label{R.6}
\end{equation}%
and the operator $\hat{H}_{j,\zeta }=\sideset{}{^{\,\lower1mm\hbox{$%
\oplus$}}}\sum_{M}\hat{H}_{j,M,\zeta }$ commutes with $\hat{J}_{1,2j,\zeta }$%
. Then the closure of the direct orthogonal sum 
\begin{equation}
\hat{H}\mathbf{=}\sideset{}{^{\,\lower1mm\hbox{$\oplus$}}}\sum_{j,\zeta }%
\hat{H}_{j,\zeta }\mathbf{=}\sideset{}{^{\,\lower1mm\hbox{$\oplus$}}}%
\sum_{j,M,\zeta }\hat{H}_{j,M,\zeta }  \label{R.7}
\end{equation}%
is a s.a. operator in the whole Hilbert space $\mathcal{H}$~\cite{19}, and $%
\hat{H}$ is a rotationally invariant extension of the rotationally invariant
symmetric operator $\hat{H}^{\left( 0\right) }$.

Conversely, any rotationally invariant s.a. extension $\hat{H}$ of the
initial operator $\hat{H}^{\left( 0\right) }$ has structure (\ref{R.7}), and
the operator $\hat{h}_{j,\zeta }=U_{j,M,\zeta }\hat{H}_{j,M,\zeta
}U_{j,M,\zeta }^{-1}$ in $\mathcal{L}^{2}(0,\infty )$ is independent of $M$
and is a s.a. extension of the symmetric operator $\hat{h}_{j,\zeta
}^{\left( 0\right) }$.\footnote{%
Roughly speaking, this means that s.a. extensions of the parts $\hat{H}%
_{j,M,\zeta }^{\left( 0\right) }$ with fixed $j$and $\zeta $ and different $%
M^{\prime }$s must be constructed ``uniformly''.}

The problem of constructing a rotationally invariant s.a. Hamiltonian $%
\hat{H}$ is thus reduced to the problem of constructing s.a. radial
Hamiltonians $\hat{h}_{j,\zeta}$.

In what follows, we operate with fixed $j$ and $\zeta$ and therefore omit
these indices for brevity. In fact, we consider the radial differential
expressions $\check{h}_{j,\zeta}\;$as a two-parameter differential
expression $\check{h}$ with the parameters $q$ and $\varkappa$ (the
parameters $j$ and $\zeta$ enter through one parameter $\varkappa$, the
parameter $m$ is considered fixed) and similarly treat the associated radial
operators $\hat{h}^{(0)}$ and $\hat{h}$\ defined in the same Hilbert space $%
\mathcal{L}^{2}(0,\infty)$.

\section{General solution of radial equations}

\label{sec3}

Later on, we need some special solutions of the differential equation 
\begin{equation*}
\check{h}F=WF,
\end{equation*}%
or, which is the same, the system of equations 
\begin{equation}
\left\{ 
\begin{array}{c}
\frac{df}{dr}+\frac{\varkappa }{r}f-(W+m+\frac{q}{r})g=0, \\ 
\frac{dg}{dr}-\frac{\varkappa }{r}g+(W-m+\frac{q}{r})f=0,%
\end{array}%
\right.  \label{1.14}
\end{equation}%
with an arbitrary complex $W$; real $W$ are denoted by $E$ and have the
conventional sense of energy. We call Eqs. (\ref{1.14}) the radial
equations. For completeness, we present the general solution of the radial
equations following the standard procedure, see, e.g., \cite{13,14}. We
first represent $f(r)$ and $g(r)$ as 
\begin{equation*}
f(r)=z^{\Upsilon }e^{-z/2}[P\left( z\right) +Q(z)]\,,\;g(r)=-i\Lambda
z^{\Upsilon }e^{-z/2}[P(z)-Q(z)]\,,
\end{equation*}%
where $\;z=-2iKr\,,\Upsilon $, $\Lambda ,$ and $K$ are some complex numbers
that are specified below. The radial equations then become the equations for
the functions $P$ and $Q$. Setting 
\begin{align}
& \Upsilon ^{2}=\varkappa ^{2}-q^{2},\;\alpha =\Upsilon -i\frac{qW}{K}%
,\;\beta =1+2\Upsilon ,  \notag \\
& W\pm m=\rho _{\pm }e^{i\varphi _{\pm }},\;0\leq \varphi _{\pm }<2\pi \,,
\label{1.15a} \\
& \Lambda =\sqrt{\frac{W-m}{W+m}}=\sqrt{\frac{\rho _{-}}{\rho _{+}}}e^{\frac{%
i}{2}(\varphi _{-}-\varphi _{+})}\,,\;K=\sqrt{W^{2}-m^{2}}=\sqrt{\rho
_{-}\rho _{+}}e^{\frac{i}{2}(\varphi _{-}+\varphi _{+})},  \notag
\end{align}%
we reduce (\ref{1.14}) to the system of equations 
\begin{align}
& z\frac{d^{2}Q}{dz^{2}}+\left( \beta -z\right) \frac{dQ}{dz}-\alpha Q=0,\, 
\notag \\
& P=-\frac{1}{\varkappa -i(qm/K)}\left( z\frac{d}{dz}+\alpha \right) Q.
\label{1.19}
\end{align}%
The first equation in (\ref{1.19}) is the confluent hypergeometric equation~%
\cite{20,21} for $Q$.

Let\footnote{%
The parameter $\Upsilon $ is defined by (\ref{1.15a}) up to a sign. A
specification of $\ \Upsilon \ $\ is a matter of convenience. In particular,
for specific values of charge, we also use a specification of $\Upsilon $
where $\Upsilon =-n/2$, this case is considered separately below.} $\Upsilon
\neq -n/2$, $n=1,2,...$, then its general solution can be represented as 
\begin{equation}
Q=A\Phi (\alpha ,\beta ;z)+B\Psi (\alpha ,\beta ;z)\,,  \label{1.22}
\end{equation}%
where $A$ and $B$ are arbitrary constants, $\Phi(\alpha,\beta;z)$ and $%
\Psi(\alpha,\beta;z)$ are the known confluent hypergeometric functions, for
their definition, see~\cite{20,21} (the function $\Phi (\alpha ,\beta ;z)$
is not defined for $\beta =0,-1,-2,...$). It follows from eqs. (\ref{1.19})
and (\ref{1.22}) that 
\begin{equation*}
P=-\frac{\alpha }{\varkappa -i(qm/K)}\left[ A\Phi (\alpha +1,\beta
;z)-B(\Upsilon +i\frac{qW}{K})\Psi \left( \alpha +1;\beta ;z\right) \right] .
\end{equation*}%
The general solution of radial equations (\ref{1.14}) for any complex $W$
and real$\,m$, $\varkappa $, and $q$ is finally given by 
\begin{align*}
& f(r)=z^{\Upsilon }e^{-z/2}\{A[\Phi (\alpha ,\beta ;z)-a_{+}\Phi (\alpha
+1,\beta ;z)]+B[\Psi (\alpha ,\beta ;z)+b\Psi (\alpha +1;\beta ;z)]\}, \\
& g(r)=i\Lambda z^{\Upsilon }e^{-z/2}\{A[\Phi (\alpha ,\beta ;z)+a_{+}\Phi
(\alpha +1,\beta ;z)]+B[\Psi (\alpha ,\beta ;z)-b\Psi (\alpha +1;\beta
;z)]\}, \\
& a_{\pm }=\frac{\pm \Upsilon K-iqW}{\varkappa K-iqm},\;b=\frac{\varkappa
K+iqm}{K}.
\end{align*}%
Taking the relation 
\begin{equation*}
\Phi (\alpha +1,\beta ;-2iKr)=e^{-2iKr}\Phi (\beta -\alpha -1,\beta ;2iKr)
\end{equation*}%
into account, see~\cite{20,21}, it is convenient to represent the general
solution of radial equations (\ref{1.14}) in the form 
\begin{equation}
F=\left( 
\begin{array}{c}
f \\ 
g%
\end{array}%
\right) =AX(r,\Upsilon ,W)+Bz^{\Upsilon }e^{-z/2}\left[ \Psi (\alpha ,\beta
;z)\vartheta _{+}-b\Psi (\alpha +1,\beta ;z)\vartheta _{-}\right] ,
\label{1.23}
\end{equation}%
where the doublets $\vartheta _{\pm }$ are 
\begin{equation*}
\vartheta _{\pm }=\left( 
\begin{array}{c}
\pm 1 \\ 
i\Lambda%
\end{array}%
\right)
\end{equation*}%
and the doublet $X$ is 
\begin{align}
& X=\frac{(mr)^{\Upsilon }}{2}\left[ \Phi _{+}(r,\Upsilon ,W)+\Phi
_{-}(r,\Upsilon ,W)\left( 
\begin{array}{cc}
0 & m+W \\ 
m-W & 0%
\end{array}%
\right) \right] u_{+},  \notag \\
& \Phi _{+}=e^{iKr}\Phi \left( \Upsilon +\frac{qW}{iK},1+2\Upsilon
;-2iKr\right) +e^{-iKr}\Phi \left( \Upsilon -\frac{qW}{iK},1+2\Upsilon
;2iKr\right) \,,  \label{1.28} \\
& \Phi _{-}=\frac{1}{iK}\left[ e^{iKr}\Phi \left( \Upsilon +\frac{qW}{iK}%
,1+2\Upsilon ;-2iKr\right) -e^{-iKr}\Phi \left( \Upsilon -\frac{qW}{iK}%
,1+2\Upsilon ;2iKr\right) \right] ,  \notag
\end{align}%
the doublet $u_{+}$ is one of doublets $u_{\pm }$ which are used below, 
\begin{equation*}
u_{\pm }=\left( 
\begin{array}{c}
1 \\ 
\frac{\varkappa \pm \Upsilon }{q}%
\end{array}%
\right) .
\end{equation*}

We now present some particular solutions of radial equations (\ref{1.14})
which are used in the following.

One of the solutions given by\ (\ref{1.23}) with $A=1,\,B=0$, and a specific
choice for $\Upsilon $ is 
\begin{equation}
U_{(1)}(r;W)=\left. X(r,\Upsilon ,W)\right| _{\Upsilon =\Upsilon _{+}},
\label{1.27}
\end{equation}%
where 
\begin{equation*}
\Upsilon _{+}=\left\{ 
\begin{array}{l}
\sqrt{\varkappa ^{2}-q^{2}},\;q\leq |\varkappa |, \\ 
\sqrt{q^{2}-\varkappa ^{2}},\;q>|\varkappa |,%
\end{array}%
\right.
\end{equation*}%
in what follows, we set $\gamma =\sqrt{\varkappa ^{2}-q^{2}}$ and$\sigma =%
\sqrt{q^{2}-\varkappa ^{2}}$. The asymptotic behavior of the doublet $%
U_{(1)}(r;W)$ at the origin is given by 
\begin{equation}
U_{(1)}(r;W)=(mr)^{\Upsilon _{+}}u_{+}+O(r^{\Upsilon _{+}+1}),\;r\rightarrow
0.  \label{1.30}
\end{equation}

In the case where $\Upsilon _{+}\neq n/2$, $n=1,2,...$, we also use another
solution 
\begin{equation}
U_{(2)}(r;W)=\left. X(r,\Upsilon ,W)\right| _{\Upsilon =-\Upsilon _{+}}
\label{1.31}
\end{equation}%
with the asymptotic behavior 
\begin{equation}
U_{(2)}(r;W)=(mr)^{-\Upsilon _{+}}u_{-}+O(r^{-\Upsilon
_{+}+1}),\;r\rightarrow 0.  \label{1.32}
\end{equation}%
For $q\neq q_{cj}=\left| \varkappa \right| =j+\frac{1}{2}$, i.e. for $%
\Upsilon _{+}\neq 0$, the solutions $U_{(1)}(r;W)$ and $U_{(2)}(r;W)$ are
linearly independent, 
\begin{equation}
\mathrm{Wr}(U_{(1)},U_{(2)})=-\frac{2\Upsilon _{+}}{q},  \label{1.33}
\end{equation}%
where $\mathrm{Wr}(F_{1},F_{2})=F_{1}i\sigma ^{2}F_{2}=f_{1}g_{2}-g_{1}f_{2}$
is the Wronskian of the doublets $F_{1}$ $=\left( 
\begin{array}{c}
f_{1} \\ 
g_{1}%
\end{array}%
\right) $and $F_{2}=\left( 
\begin{array}{c}
f_{2} \\ 
g_{2}%
\end{array}%
\right) $.

It follows from the standard representation for the function $\Phi $ that
for real $\Upsilon \,(\Upsilon \neq -n/2)$, the functions $\Phi _{+}$ and $%
\Phi _{-}$ in (\ref{1.28}) are real-entire functions of $W$, i.e., they are
entire in $W$ and real for real $W=E$. It then follows from representations (%
\ref{1.27}) and (\ref{1.31}) that the respective doublets $U_{(1)}(r;W)$ and 
$U_{(2)}(r;W)$ are also real-entire functions of $W$ for real $\Upsilon
_{+}=\gamma $. If $\Upsilon _{+}$ is pure imaginary, $\Upsilon _{+}=i\sigma $%
, then $U_{(1)}(r;W)$ and $U_{(2)}(r;W)$ are entire in $W$ and complex
conjugate for real $W=E$, $\overline{U_{(1)}(r;E)}=U_{(2)}(r;E)$.

Another useful solution nontrivial for $\Upsilon _{+}\neq n/2$, $n=1,2,...,$
is given by (\ref{1.23}) with $A=0$ and a special choice for $B$: 
\begin{align}
& V_{\left( 1\right) }(r;W)=B(W)(mr)^{\Upsilon _{+}}e^{iKr}\left[ \Psi
(\alpha ,\beta ;z)\vartheta _{+}-b\Psi \left( \alpha +1;\beta ;z\right)
\vartheta _{-}\right] ,  \notag \\
& B(W)=\frac{\Gamma (-\Upsilon _{+}+qW/iK)}{\Gamma (-2\Upsilon _{+})(1-a_{+})%
}\equiv \frac{1}{\Gamma (-2\Upsilon _{+})\tilde{B}(W)}.  \label{1.34}
\end{align}%
As any solution, $V_{\left( 1\right) }$ is a special linear combination of $%
U_{(1)}$ and $U_{(2)}$, 
\begin{equation}
V_{(1)}(r;W)=U_{(1)}(r;W)+\frac{q}{2\Upsilon _{+}}\omega (W)U_{(2)}(r;W),
\label{1.35}
\end{equation}%
where 
\begin{align}
& \omega (W)=-\mathrm{Wr}(U_{(1)},V_{\left( 1\right) })=  \notag \\
& \,=\frac{2\Upsilon _{+}\Gamma (2\Upsilon _{+})\Gamma (-\Upsilon
_{+}+qW/iK)(1-a_{-})(2e^{-i\pi /2}K/m)^{-2\Upsilon _{+}}}{q\Gamma
(-2\Upsilon _{+})\Gamma (\Upsilon _{+}+qW/iK)(1-a_{+})}\equiv \frac{\tilde{%
\omega}(W)}{\Gamma (-2\Upsilon _{+})}.  \label{1.36}
\end{align}%
We note that if $\Im W>0$ and $r\rightarrow \infty $, the doublet $%
U_{(1)}(r;W)$ increases exponentially while $V_{(1)}(r;W)$ decreases
exponentially (with a polynomial accuracy).

The doublets $U_{(2)}$ and $V_{(1)}$\ are not solutions linearly independent
of $U_{(1)}$ at the points $\Upsilon _{+}=\gamma =n/2$ where $U_{(2)\text{ }%
} $ is not defined while $V_{(1)}$ vanishes. We need their analogues defined
at these points and having all the required properties, in particular,
real-entirety\ in\ $W.$ Unfortunately, we can construct such solutions only
in some neighborhood of fixed$\;n$. The corresponding solutions are labelled
by the index $n$.

According to (\ref{1.34}), the doublet $V_{(1)}$ tends to zero like $%
1/\Gamma (-2\gamma )$ as $\gamma \rightarrow n/2$. It then follows from (\ref%
{1.35}) that the doublet $U_{(2)}$ has a singularity of the form $\Gamma
(-2\gamma )$ at the point $\gamma =n/2$ and can be represented in a
neighborhood of this point as 
\begin{equation}
U_{(2)}(r;W)=\Gamma (-2\gamma )A_{n}(W)U_{(1)}(r;W)+U_{n(2)}(r;W),
\label{1.37}
\end{equation}
where 
\begin{equation}
A_{n}(W)=\left( -\frac{2\gamma }{q}\frac{1}{\tilde{\omega}(W)}\right)
_{\gamma =n/2}  \label{1.38}
\end{equation}
and $U_{n(2)}(r;W)$ has a finite limit as $\gamma\rightarrow n/2$ and
evidently satisfies radial equations (\ref{1.14}). A direct calculation%
\footnote{%
With the use of the equality $\Gamma(w+1)=w\Gamma(w)$.} shows that $A_{n}(W)$
is a polynomial in $W$ with real coefficients, and because $U_{(1)}(r,W)$
and $U_{(2)}(r,W)$ are real entire in $W$, the doublet $U_{n(2)}(r;W)$ is
also real entire. We thus obtain that the doublet $U_{n(2)} $ defined by 
\begin{equation}
U_{n(2)}(r;W)=U_{(2)}(r;W)-\Gamma (-2\gamma )A_{n}(W)U_{(1)}(r;W)
\label{1.38a}
\end{equation}
and satisfying the condition 
\begin{equation*}
U_{n(2)}(r;W)=(mr)^{-\gamma }u_{-}+O(r^{-\gamma +1}),\;r\rightarrow 0,
\end{equation*}
is a solution of radial equations which is well-defined in some neighborhood
of the point $\gamma=n/2$, the point itself included. This solution is
linear-independent of $U_{(1)}$, $\mathrm{Wr}\,(U_{(1)},U_{n(2)})=-2\gamma
/q $, and is real entire in $W$. According to relations (\ref{1.35})-(\ref%
{1.37}), the doublet $V_{(1)}$ is represented in a neighborhood of the point 
$\gamma =n/2$ in terms of the finite doublets $U_{(1)}$ and $U_{n(2)}$ as 
\begin{equation*}
V_{(1)}(r;W)=[1+\frac{q}{2\gamma }\tilde{\omega}(W)A_{n}(W)]U_{(1)}(r;W)+ 
\frac{q}{2\gamma }\omega (W)U_{n(2)}(r;W),
\end{equation*}
where according to (\ref{1.38}) and (\ref{1.36}), the factors $1+\frac{q}{%
2\gamma }\tilde{\omega}(W)A_{n}(W)$ and $\omega (W)\;$tend to zero like $%
1/\Gamma(-2\gamma)$ as $\gamma\rightarrow n/2$. This allows introducing the
doublet 
\begin{equation}
V_{n(1)}(r;W)=\frac{1}{1+\frac{q}{2\gamma }\tilde{\omega}(W)A_{n}(W)}%
V_{(1)}(r;W)=U_{(1)}(r;W)+\frac{q}{2\gamma}\omega_{n}(W)U_{n(2)}(r;W),
\label{eq.27a}
\end{equation}
where 
\begin{equation}
\omega _{n}(W)=\frac{\omega (W)}{1+\frac{q}{2\gamma }\tilde{\omega}%
(W)A_{n}(W)}=\frac{\tilde{\omega}(W)}{\Gamma (-2\gamma )[1+\frac{q}{2\gamma }%
\tilde{\omega}(W)A_{n}(W)]},  \label{1.38b}
\end{equation}%
which is evidently a solution of the radial equations well-defined in some
neighborhood of the point $\gamma =n/2$, the point itself included, and
exponentially decreasing as $r\rightarrow \infty $. The function $\omega
_{n}(W)$ is also well defined in some neighborhood of the point $\gamma =n/2$
and at the point itself.We point out that the useful relations 
\begin{align}
& \frac{1}{\omega _{n}(W)}V_{n(1)}(r;W)=\frac{1}{\omega (W)}V_{(1)}(r;W),
\label{1.39a} \\
& \frac{1}{\omega _{n}(W)}=\frac{q}{2\gamma }\Gamma (-2\gamma )A_{n}(W)+%
\frac{1}{\omega (W)}  \label{1.39b}
\end{align}%
hold; strictly speaking, they are meaningful for $\gamma \neq n/2$ and show
that the r.h.s.'s are continuous in $\gamma $ at the points $\gamma =n/2$.

The doublets $U_{n(2)}(r;W)$\ and $V_{n(1)}(r;W)$ are the required analogues
of$\;$the doublets\ $U_{(2)}(r;W)$\ and $V_{(1)}(r;W)$ defined in the
neighborhood of the point $\gamma =n/2$ and at the point itself.

It remains to consider the special case of $q=q_{cj}=j+1/2$, or $\Upsilon =0$%
, where the doublets $U_{(1)}$ and $U_{(2)}$ coincide while $V_{(1)}$
vanishes. Let $U_{(1)}(r;W|\gamma ),$\ $U_{(2)}(r;W|\gamma )$, and $%
V_{(1)}(r;W|\gamma )$ denote $U_{(1)}$, $U_{(2)}$, and $V_{(1)}$ with $%
\gamma \neq 0$. Differentiating radial equations (\ref{1.14}) for $U_{(1)}$
with respect to $\gamma $ at $\gamma =0$, we can easily verify that the
doublet 
\begin{equation*}
\left. \frac{\partial U_{(1)}(r;W)}{\partial \gamma }\right| _{\gamma
=0}=\lim_{\gamma \rightarrow 0}\frac{U_{(1)}(r;W|\gamma )-U_{(2)}(r;W|\gamma
)}{2\gamma }
\end{equation*}%
is a solution of these equations with $\gamma =0$. For two linearly
independent solutions of radial equations (\ref{1.14}) with $\gamma =0$, we
choose 
\begin{align}
& U_{(1)}(r;W)=U_{(1)}(r;W|0)\,,  \notag \\
& U_{(1)}(r;W)=u_{+}+O(r),\;r\rightarrow 0,  \label{1.40}
\end{align}%
and 
\begin{align}
& U_{(2)}^{(0)}(r;W)=\frac{\partial U_{(1)}(r;W|0)}{\partial \gamma }-\frac{%
\zeta }{q_{cj}}U_{(1)}(r;W|0),  \notag \\
& U_{(2)}^{(0)}(r;W)=u_{-}^{(0)}(r)+O(r\ln r),\;r\rightarrow 0,  \label{1.41}
\end{align}%
where $u_{+}$ and $u_{-}^{(0)}(r)$ are 
\begin{equation}
u_{+}=\left( 
\begin{array}{c}
1 \\ 
\zeta%
\end{array}%
\right) ,\;\;u_{-}^{(0)}(r)=\left( 
\begin{array}{c}
\ln (mr)-\frac{\zeta }{q_{cj}} \\ 
\zeta \ln (mr)%
\end{array}%
\right) .  \label{1.42}
\end{equation}%
The Wronskian of these solutions is 
\begin{equation}
\mathrm{Wr}(U_{(1)},U_{(2)}^{(0)})=\frac{1}{q_{cj}}.  \label{1.42a}
\end{equation}%
The both doublets $U_{(1)}$ and $U_{(2)}^{(0)}$ are real entire in $W$.

As \ an analogue of $V_{(1)}$ in the case of $\gamma =0$, we take the
doublet 
\begin{eqnarray}
&&V_{(1)}^{(0)}(r;W)=\lim_{\gamma \rightarrow 0}[-\Gamma (-2\gamma
)V_{(1)}(r;W|\gamma )]=  \label{1.43} \\
&=&-\frac{\Gamma (\alpha )}{1-a}e^{iKr}\left[ \Psi (\alpha ,1;-2iKr)+b\Psi
\left( \alpha +1,1;-2iKr\right) \sigma ^{3}\right] \left( 
\begin{array}{c}
1 \\ 
i\Lambda%
\end{array}%
\right) ,  \notag
\end{eqnarray}

where%
\begin{equation*}
\alpha =\frac{q_{cj}W}{iK},\;a=\frac{W}{m+i\zeta K},\;b=q_{cj}\frac{\zeta
K+im}{K}.
\end{equation*}

Its representation in terms of $U_{(1)}$ and $U_{(2)}^{(0)}$ is given by 
\begin{align}
& V_{(1)}^{(0)}(r;W)=U_{(2)}^{(0)}(r;W)+q_{cj}\omega ^{(0)}(W)U_{(1)}(r;W), 
\notag \\
& \omega ^{(0)}(W)=-\mathrm{Wr}(U_{(2)}^{(0)}(r;W),V_{(1)}^{(0)}(r;W))=
\label{1.44} \\
& =\frac{1}{q_{cj}}\left[ \ln (2e^{-i\pi /2}K/m)+\psi (-iq_{cj}W/K)+\frac{%
\zeta (W-m)+iK}{2q_{cj}W}-2\psi (1)\right] ,  \notag
\end{align}%
where $\psi \;$is\ a symbol of the logarithmic derivative of the $\Gamma $%
-function. We note that if $\Im W>0,$ the doublet $V_{(1)}^{(0)}(r;W)$ is
square integrable, $V_{(1)}^{(0)}(r;W)\in \mathcal{L}^{2}(0,\infty )$, and
exponentially decreases as $r\rightarrow \infty $.

\section{Self-adjoint radial Hamiltonians for different regions of charge $q$%
}

\label{sec4}

\subsection{Generalities}

\label{sec4.1}

In this section, we construct a s.a. radial Hamiltonian\footnote{%
We omit indices in the notation of $\hat{h}$, $\hat{h}^{(0)}$, and $\check{h}
$, see the end of Sec.~\ref{sec2}.} $\hat{h}$ in the Hilbert space $\mathcal{%
L}^{2}(0,\infty )$ of doublets as a s.a. extension of symmetric radial
operator $\hat{h}^{(0)}$ (\ref{R.3}) associated with the radial differential
expression $\check{h}\;$(\ref{R.4})\emph{\ }and analyze its spectral
properties.

The extension procedure includes the following steps~\cite{22}:

1) evaluating the adjoint operator $\hat{h}^{\ast}=\left( \hat{h}^{\left(
0\right) }\right) ^{+}\,$and estimating its asymmetricity in terms of
(asymptotic) boundary values of doublets belonging to the domain $D_{\ast}$
of $\hat{h}^{\ast}$.

2) constructing s.a. extensions $\hat{h}$ of $\hat{h}^{\left( 0\right) }$ as
s.a. restrictions of the adjoint $\hat{h}^{\ast }$ specified by some
(asymptotic) s.a. boundary conditions at the origin\footnote{%
We note that this method of constructing $\hat{h}$ allows avoiding \ an
evaluation of the deficient \ subspaces \ and deficiency indices of $\hat{h}%
^{(0)}$, the latters are determined by passing.}.

It turns out\footnote{%
Actually, this becomes clear at the first step.} that the result crucially
depends on the value of the charge $q$: different regions of the charge are
assigned different s.a. radial Hamiltonians in the sense that they are
specified by completely different types of (asymptotic) s.a. boundary
conditions. What is more, for sufficiently large charges, a s.a. radial
Hamiltonian is defined non-uniquely, such that there is a one-parameter
family of s.a. Hamiltonians for fixed $\varkappa $ and$\;q$. Therefore, our
exposition is naturally divided into subsections related to the
corresponding regions of the charge; actually, there are four of them.

For each region, we perform a full spectral analysis of the obtained s.a.
Hamiltonians, in particular, we find their spectra and (generalized)
eigenfunctions. The analysis is based on the Krein method of guiding
functionals and includes the following steps:

i) constructing the guiding functional,

ii) evaluating the resolvent,

iii) evaluating the spectral function,

iv) constructing the so-called inversion formulas\ that are mathematically
rigorous formulas for the Fourier expansion of wave functions with respect
to the (generalized) eigenfunctions of the s.a. Hamiltonian.

We compare this analysis with heuristic physical considerations based, in
particular, on the rule of ``normalization to $\delta $-function'' for
eigenfunctions of the continuous spectrum.

Our first task is constructing a s.a. radial operator $\hat{h}$ in
accordance with the above scheme, which mainly reduces to indicating its
domain $D_{h}$, $D_{h^{(0)}}\subset D_{h}\subseteq D_{\ast }$. It can happen
that such a domain is non-unique, and it really is for some values of number
parameters in $\check{h}$.

The domain for a s.a. differential operator on an interval of the real axis
is conventionally specified by the so-called s.a. boundary conditions at the
ends of the interval for the functions belonging to\emph{\ }the domain. Our
task is to indicate these conditions for the doublets $F$ $\in $ $D_{h}$ at
the boundaries $r=0$ and $r=\infty $ which are the so-called singular ends
of the differential expression $\check{h}$ (\ref{R.4}), see~\cite{18}. The
singularity of the left end $r=0$ is due to the nonintegrability of the free
terms (the coefficient functions without derivatives) in $\check{h}$. In the
case of singular ends, there is no universal explicit method for formulating
s.a. boundary conditions. In our case where the coefficient function in
front of the derivative $d/dr$ is independent of $r,$ while the free terms
are bounded as $r\rightarrow \infty $, the problem of s.a. boundary
conditions is related only to the left end $r=0$.

We begin with the adjoint $\hat{h}^{\ast }$ of the initial symmetric
operator $\hat{h}^{\left( 0\right) }$. Using the known distribution theory
arguments (or extending the known results for scalar differential operators~%
\cite{18} to the matrix differential operators), we can easily verify that
the adjoint operator $\hat{h}^{\ast }$ is given by 
\begin{equation}
\hat{h}^{\ast }:\left\{ 
\begin{array}{l}
D_{h^{\ast }}=D_{\ast }=\left\{ 
\begin{array}{l}
F:F\ \text{\textrm{are absolutely\ continuous in }}\left( 0,\infty \right) ,
\\ 
F,\check{h}F=G\in \mathcal{L}^{2}(0,\infty ),%
\end{array}%
\right\} , \\ 
\hat{h}^{\ast }F\left( r\right) =\check{h}F\left( r\right) ,%
\end{array}%
\right.  \label{2.6}
\end{equation}%
i.e., the adjoint $\hat{h}^{\ast }$ is associated with the same differential
expression $\check{h}$, but defined on a more wide domain $D_{\ast }$, $%
D_{h^{(0)}}\subset D_{\ast }$, which is the so-called natural domain for the
differential expression $\check{h}$. Because the coefficient functions of
the differential expression $\check{h}$ are real, it follows that the
deficiency indices of the initial symmetric operator $\hat{h}^{\left(
0\right) }$are equal and therefore s.a. extensions of $\hat{h}^{\left(
0\right) }$ do exist for any values of parameters $\varkappa $ and $q$.

It is convenient to introduce a quadratic asymmetry form $\Delta _{\ast }$
for $\hat{h}^{\ast }$ by 
\begin{align}
& \Delta _{\ast }\left( F\right) =\left( F,\hat{h}^{\ast }F\right) -\left( 
\hat{h}^{\ast }F,F\right) =2i\Im\left( F,\hat{h}^{\ast }F\right) =  \notag \\
& \,=\int_{0}^{\infty }drF^{+}\left( r\right) \left( \check{h}F\right)
\left( r\right) -\int_{0}^{\infty }dr\left( \check{h}F\right) ^{+}\left(
r\right) \,F\left( r\right) .  \label{2.7}
\end{align}%
The quantity $\Delta _{\ast }\left( F\right) $ is evidently pure imaginary%
\footnote{%
The quadratic asymmetry form $\Delta _{\ast }$ is a restriction to the
diagonal of the sesquilinear anti-Hermitian asymmetry form $\omega_{\ast}$
defined by 
\begin{eqnarray*}
&&\omega _{\ast }(F_{1},F_{2})=(F_{1},\hat{h}_{2}^{\ast }F_{2})-(\hat{h}%
_{1}^{\ast }F_{1},F_{2})= \\
&=&\int_{0}^{\infty }drF_{1}^{+}\left( r\right) \left( \check{h}F_{2}\right)
\left( r\right) -\int_{0}^{\infty }dr\left( \check{h}F_{1}\right) ^{+}\left(
r\right) \,F_{2}\left( r\right) ,
\end{eqnarray*}%
the forms $\Delta _{\ast }$ and $\omega _{\ast }$define each other~\cite{22}.%
}. The form $\Delta _{\ast }$ yields a measure of the asymmetricity of the
operator $\hat{h}^{\ast }$, it shows to what extent the operator $\hat{h}%
^{\ast }$ is nonsymmetric. If $\Delta _{\ast }\equiv 0$, the operator $%
\hat{h}^{\ast }$ is symmetric and therefore s.a..This also means that $%
\hat{h}^{\left( 0\right) }$ is essentially s.a. and its unique s.a.
extension is its closure, $\hat{h}=\overline{\hat{h}^{\left( 0\right) }}$,
which coincides with the adjoint, $\hat{h}=\hat{h}^{\ast }=\left( \hat{h}%
\right) ^{+}$. If $\Delta _{\ast }\neq 0$, a s.a. operator $\hat{h}=\left( 
\hat{h}\right) ^{+}$ is constructed as a restriction of the operator $\hat{h}%
^{\ast }$ to the domain $D_{h}\subset D_{\ast }$ such that the restriction
of $\Delta _{\ast }$ to $D_{h}$ vanishes and $D_{h}$ is a maximum domain 
\footnote{%
I.e., a domain that does not allow further extending with the condition $%
\Delta _{\ast }\equiv 0$ conserved.}~\cite{22}.

Using integrating by parts in r.h.s. in (\ref{2.7}), it is easy to verify
that $\Delta _{\ast }$ is represented as 
\begin{equation}
\Delta _{\ast }\left( F\right) =\left[ F\right] \left( \infty \right) -\left[
F\right] \left( 0\right) \,,  \label{2.8}
\end{equation}%
where 
\begin{equation*}
\left[ F\right] \left( \infty \right) =\lim_{r\rightarrow \infty }\left[ F%
\right] \left( r\right) \ \mathrm{and}\ \left[ F\right] \left( 0\right)
=\lim_{r\rightarrow 0}\left[ F\right] \left( r\right)
\end{equation*}%
are the corresponding boundary values of the quadratic local form $\left[ F%
\right] \left( r\right) $ defined by\footnote{%
The quadratic local form $[F](r)$ is a restriction to the diagonal of yhe
sesquilinear anti-Hermitian local form \newline
$[F_1,F_2](r)=-iF\sigma^2F_2(r)=-[\overline{f_1(r)}g_2(r) -\overline{g_1(r)}%
f_2(r)]$. These two forms define each other.} 
\begin{equation*}
\left[ F\right] \left( r\right) =-iF^{+}(r)\sigma ^{2}F(r)=-[\overline{f(r)}%
g(r)-\overline{g(r)}f(r)]=-2i\Im\overline{f(r)}g(r).
\end{equation*}%
These boundary values certainly exist because of the existence of the
integrals in r.h.s. in (\ref{2.7}); for brevity, we call them the boundary
forms at infinity and at the origin respectively. The asymmetry form $\Delta
_{\ast }$ is thus simply determined by the boundary forms\footnote{%
The representation (\ref{2.8}) is a particular case of the so-called
Lagrange identity in the integral form.} $\left[ F\right] \left( \infty
\right) $ and $\left[ F\right] \left( 0\right) $.

We prove that for any $F\in D_{\ast }$, we have 
\begin{equation*}
\lim_{r\rightarrow \infty }F\left( r\right) =0.
\end{equation*}%
We first note that $F\in D_{\ast }$ implies that $\check{h}F=G$ is square
integrable together with $F$. It in turn follows that 
\begin{equation*}
\frac{dF(r)}{dr}=\left( -\frac{\varkappa }{r}\sigma ^{3}+i\frac{q}{r}\sigma
^{2}+m\sigma ^{1}\right) F(r)+i\sigma ^{2}G(r)
\end{equation*}%
is square integrable at infinity \footnote{%
A doublet $F(r)$ is square integrable at infinity if $\int_{R}^{\infty
}drF^{+}(r)F(r)<\infty $ for sufficiently large $R$.}. It now remains to
refer to the assertion that if an absolutely continuous $F(r)$ is square
integrable at infinity together with its derivative $dF(r)/dr$, then $%
F(r)\rightarrow 0$ as $r\rightarrow \infty $; this assertion is an evident
generalization of a similar assertion for scalar functions. Therefore, the
boundary form at infinity is identically zero: for any$\,F\in D_{\ast }$, we
have 
\begin{equation*}
\left[ F\right] \left( \infty \right) =0,
\end{equation*}%
and the asymmetry form $\Delta _{\ast }$ is determined by the boundary form
at the origin: for any $F\in D_{\ast }$, we have 
\begin{equation}
\Delta _{\ast }\left( F\right) =-\left[ F\right] \left( 0\right) =\left. (%
\overline{f}g-\overline{g}f)\right| _{r=0}.  \label{2.12}
\end{equation}%
For evaluating this boundary form, we have to find the asymptotic behavior
of the doublets $F\in D_{\ast }$ at the origin. It turns out that the
doublets $F$ can tend to zero, be finite, or infinite, or even have no limit
as $r\rightarrow 0$ depending on the values of the parameters $\varkappa $
and $q$: at fixed $j$, we must distinguish four regions of the charge $q$
that are defined by the two characteristic values $q_{uj}$ and $q_{cj}$ ($%
q_{uj}<q_{cj}$) of the charge,

\begin{align*}
& q_{uj}=\sqrt{\varkappa^{2}-\frac{1}{4}}=\sqrt{j\left( j+1\right) }%
\;\Longleftrightarrow \\
& \Upsilon=\gamma=\frac{1}{2}\;\Longleftrightarrow\;Z_{uj}=137\sqrt{j\left(
j+1\right) }, \\
& q_{cj}=\left| \varkappa\right| =j+\frac{1}{2}\;\Longleftrightarrow \emph{\;%
}\Upsilon=0\;\Longleftrightarrow\emph{\;}Z_{cj}=137(j+1/2).
\end{align*}

The evaluation of the asymptotic behavior of $F\in D_{\ast }$ at the origin
is based on the following observation. According to definition (\ref{2.6}),
the doublets $F\in D_{\ast }$ can be considered as square-integrable
solutions of the inhomogeneous differential equation 
\begin{equation}
\check{h}F\left( r\right) =\left( -i\sigma ^{2}\frac{d}{dr}+\frac{\varkappa 
}{r}\sigma ^{1}-\frac{q}{r}+m\sigma ^{3}\right) F\left( r\right) =G\left(
r\right) \,,  \label{2.14}
\end{equation}%
with the r.h.s. $G\mathcal{\ }$belonging to\ $\mathcal{L}^{2}(0,\infty )$
and therefore locally integrable, which allows applying the general theory
of differential equations (see e.g.~\cite{18}) to equation (\ref{2.14}). For
estimating the asymptotic behavior of $F\left( r\right) $ at the origin, it
is convenient to represent (\ref{2.14}) as 
\begin{equation}
\check{h}_{-}F\left( r\right) =G_{-}\left( r\right) \,,  \label{2.4}
\end{equation}%
where 
\begin{equation*}
\check{h}_{-}=-i\sigma ^{2}\partial _{r}+\frac{\varkappa }{r}\sigma ^{1}-%
\frac{q}{r}\,,\ \ G_{-}\left( r\right) =G\left( r\right) -m\sigma
^{3}F\left( r\right) \in \mathcal{L}^{2}(0,\infty )\,.
\end{equation*}%
Let $U_{1\;}$and $U_{2}$ be linearly independent solutions of the \
homogeneous differential equation $\check{h}_{-}U=0$, 
\begin{align}
& U_{1}(r)=(mr)^{\Upsilon _{+}}u_{+}\,,\;q>0,  \notag \\
& U_{2}(r)=\left\{ 
\begin{array}{l}
(mr)^{-\Upsilon _{+}}u_{-}\,,\;q>0,\;q\neq q_{cj}\;(\Upsilon _{+}\neq 0)\,,
\\ 
u_{-}^{(0)}(r)=\left( 
\begin{array}{c}
\ln (mr)-\zeta /q_{cj} \\ 
\zeta \ln (mr)%
\end{array}%
\right) ,\;q=q_{cj}\;(\Upsilon _{+}=0)\,.%
\end{array}%
\right.  \label{2.15}
\end{align}%
Any solution $F\left( r\right) $ of inhomogeneous differential equation (\ref%
{2.4}) can be represented as 
\begin{equation}
F(r)=c_{1}U_{1}(r)+c_{2}U_{2}(r)+I_{1}(r)+I_{2}(r)\,,  \label{2.16}
\end{equation}%
where $c_{1}$ and $\,c_{2}$\ are some constants and 
\begin{align}
I_{1}(r)& =\left\{ 
\begin{array}{l}
\frac{q}{2\Upsilon _{+}}\int_{r}^{r_{0}}[U_{1}(r)\otimes
U_{2}(y)]G_{-}(y)dy\,,\;0<q\leq q_{uj}\,, \\ 
-\frac{q}{2\Upsilon _{+}}\int_{0}^{r}[U_{1}(r)\otimes
U_{2}(y)]G_{-}(y)dy\,,\;q>q_{uj}\,,\;q\neq q_{cj}\,, \\ 
q_{cj}\int_{0}^{r}[U_{1}(r)\otimes U_{2}(y)]G_{-}(y)dy\,,\;q=q_{cj}\,,%
\end{array}%
\right.  \notag \\
I_{2}(r)& =\left\{ 
\begin{array}{l}
\frac{q}{2\Upsilon _{+}}\int_{0}^{r}[U_{2}(r)\otimes
U_{1}(y)]G_{-}(y)dy\,,\;q>0,\;q\neq q_{cj}\,, \\ 
-q_{cj}\int_{0}^{r}[U_{2}(r)\otimes U_{1}(y)]G_{-}(y)dy\,,\;q=q_{cj}\,,%
\end{array}%
\right.  \label{2.17}
\end{align}%
where $\otimes $ is the symbol of tensor product, such that $%
[U_{1}(r)\otimes U_{2}(y)]$ is a $2\times 2$ matrix , and $r_{0}>0$ is a
constant. It turns out that the boundary form $[F]\left( 0\right) $ is
determined by the two first terms in r.h.s. in representation (\ref{2.16})
and essentially depends on the parameter $\Upsilon $.

\subsection{First noncritical region}

\label{sec4.2}

The first noncritical region of the charge is defined by the condition 
\begin{equation*}
0<q\leq q_{uj}\,\Longleftrightarrow \text{ }\Upsilon _{+}=\gamma \geq \frac{1%
}{2}\,.
\end{equation*}

\subsubsection{Self-adjoint radial Hamiltonians}

\label{sec4.2.1}

The representation given by formulas (\ref{2.15}) -- (\ref{2.17}) allows
evaluating the asymptotic behavior of $F\in D_{\ast }$ at the origin.
According to (\ref{2.15}), the doublet $U_{1}(r)\sim r^{\gamma }$ is square
integrable at the origin, whereas the doublet$\ U_{2}(r)\sim r^{-\gamma }$
is not. Using the Cauchy-Bounjakowsky inequality for estimating the
integrals $I_{1}(r)$ and $I_{2}(r)$ (\ref{2.17}), we find 
\begin{equation}
I_{1}(r)=O(r^{1/2})\,,\;I_{2}(r)=O(r^{1/2}),\;r\rightarrow 0.  \label{2.19}
\end{equation}%
It follows that for $F\left( r\right) $ to belong to the space $\mathcal{L}%
^{2}(0,\infty )$, it is necessary that the coefficient $c_{2}$ in front of $%
U_{2}\left( r\right) $ in (\ref{2.16}) be zero, $c_{2}=0$, which yields 
\begin{equation}
F(r)=c_{1}U_{1}(r)+I_{1}(r)+I_{2}(r)=O(r^{1/2})\,\rightarrow
0,\;r\rightarrow 0,  \label{2.19a}
\end{equation}%
whence it follows that for any $F\in D_{\ast }$, we have 
\begin{equation*}
\left[ F\right] \left( 0\right) =0.
\end{equation*}%
This means that in the first noncritical charge region, $0<q\leq q_{uj}$, or 
$\gamma \geq 1/2$, the operator $\hat{h}^{\ast }=\hat{h}$ is s.a. and it is
a unique s.a. operator associated with s.a. differential expression $\check{h%
}$ (\ref{R.4})\footnote{%
We point out a particular corollary: the deficiency indices of the initial
symmetric operator $\check{h}^{(0)}$ are $(0,0)$ in the charge region $%
0<q\leq q_{uj}$\ $(\gamma \geq 1/2)$.
\par
{}}, \thinspace such that 
\begin{equation}
\hat{h}:\left\{ 
\begin{array}{l}
D_{h}=\left\{ 
\begin{array}{l}
F:F\text{\textrm{\ are absolutely\ continuous in\ }}\left( 0,\infty \right) ,
\\ 
F,\,\check{h}F\in \mathcal{L}^{2}(0,\infty ),%
\end{array}%
\right\} \\ 
\hat{h}F\left( r\right) =\check{h}F\left( r\right) .%
\end{array}%
\right.  \label{2.21}
\end{equation}

We note that this result actually justifies the standard treatment of the
Dirac Hamiltonian \ with $q\leq \sqrt{3}/2$, or $Z\leq 118$, in the physical
literature where the natural domain for $\check{h}$ is implicitly assumed%
\footnote{%
The uniqueness of the Hamiltonian also implies that the notion of $\delta $
potential for a relativistic Dirac particle cannot be introduced, which
maybe manifests the nonrenormalizability of the four-fermion interaction.}.

We now proceed to the spectral analysis of the obtained Hamiltonian in
accordance with the scheme described in Subsec~\ref{sec4.1}. A necessary
short information on each item of this scheme is given in Appendix~\ref{appA}%
.

\subsubsection{Spectral analysis}

\label{sec4.2.2}

\paragraph{Guiding functional}

In accordance with the requirements in Appendix~\ref{appA}, for the doublet $%
U$ defining guiding functional $\Phi (F;W)$ (\ref{A1.0}), we choose the
doublet $U_{(1)}(r;W)$ given by(\ref{1.27}) -- (\ref{1.30}), 
\begin{equation*}
U(r;W)=U_{(1)}(r;W),
\end{equation*}
the doublet $U(r;W)$ is real-entire,see Sec.~\ref{sec3}.

For $\mathcal{D}$,$\;$we choose the set of doublets $F(r)\in D_{h}=D_{\ast}$
with a compact support. It is evident that $\mathcal{D}$ is dense in $%
\mathcal{L}^{2}(0,\infty)$.

The guiding functional $\Phi $ with these $U$ and $\mathcal{D}$ is simple,
i.e., satisfies the properties 1)--3) presented in Appendix~\ref{appA}. The
property 1) is evident, the property 3) is easily verified by integrating by
parts, and\ it remains to verify the property 2): the equation 
\begin{equation*}
(\check{h}-E_{0})\Psi (r)=F_{0}(r),
\end{equation*}%
where$\;F_{0}\in $ $\mathcal{D\;}$and satisfies the condition 
\begin{equation*}
\Phi (F_{0};E_{0})=\int_{0}^{\infty }U(r;E_{0})F_{0}(r)dr=0,
\end{equation*}%
has a solution belonging to $\mathcal{D}$.

At this point, our exposition is divided into two parts because it is
convenient to consider the cases of $\gamma\neq n/2$ and $\gamma=n/2$, $%
n=1,2,...$, separately. We first consider the case of $\gamma\neq n/2$,
after which the extension of the obtained results to the case of $%
\gamma=n/2\;$becomes\ evident.

In the case of $\gamma \neq n/2$, any solution of the inhomogeneous equation
allows the representation 
\begin{align}
& \Psi (r)=c_{1}U(r;E_{0})+c_{2}U_{(2)}(r;E_{0})+\frac{q}{2\gamma }%
\int_{r}^{\infty }[U(r;E_{0})\otimes U_{(2)}(y;E_{0})]F_{0}(y)dy+  \notag \\
& +\frac{q}{2\gamma }\int_{0}^{r}[U_{(2)}(r;E_{0})\otimes
U(y;E_{0})]F_{0}(y)dy,  \label{2.1.1}
\end{align}%
where $U_{(2)}(r;W)$ is given by (\ref{1.31}) -- (\ref{1.33}). This
representation is a copy of representation (\ref{2.15}) -- (\ref{2.17}) for
a solution of Eq. (\ref{2.4}), where we can take $r_{0}=\infty $ because of
the compactness of the support of $F_{0}$. The integral terms in this
representation have a compact support: if $\mathrm{supp}F_{0}\subset \lbrack
a,b],\ 0\leq a<b<\infty $, they vanish for $r>b$. Choosing $c_{1}=c_{2}=0$,
we obtain a particular solution $\Psi $ with a compact support that has the
form 
\begin{align}
& \Psi (r)=\frac{q}{2\gamma }\int_{r}^{\infty }[U(r;E_{0})\otimes
U_{(2)}(y;E_{0})]F_{0}(y)dy+  \notag \\
& +\frac{q}{2\gamma }\int_{0}^{r}[U_{(2)}(r;E_{0})\otimes
U(y;E_{0})]F_{0}(y)dy.  \label{2.1.2}
\end{align}%
Taking the asymptotic behavior of the doublets $U=$ $U_{(1)}$ and $U_{(2)}$
at the origin (see (\ref{1.30}), (\ref{1.32})), and estimate (\ref{2.19a})
for $F_{0}$ into account, we find that the asymptotic behavior of this
solution as $r\rightarrow 0$ is of the form $\Psi (r)=O(r^{\delta })$, $%
\delta =\min (\gamma ,3/2)$, whence it follows that $\Psi \in \mathcal{D}$.

\paragraph{Green's function}

To find Green's function $G(r,r^{\prime };W)$, $\Im W\neq 0$, of the s.a.
operator $\hat{h}$ associated with the s.a differential expression $\check{h}
$ is to represent a unique solution $\Psi (r)\in D_{h}$ of the differential
equation 
\begin{equation}
(\check{h}-W)\Psi (r)=F(r)  \label{2.2.0a}
\end{equation}%
with any $F(r)\in \mathcal{L}^{2}(0,\infty )\;$in\ the\ integral form 
\begin{equation}
\Psi (r)=\int_{0}^{\infty }G(r,r^{\prime };W)F(r^{\prime })dr^{\prime }.
\label{2.2.0b}
\end{equation}%
For our purposes, it is sufficient to consider the case of $\Im W>0$. As any
solution, $\Psi \;$allows the representation

\begin{align*}
& \Psi (r)=c_{1}U(r;W)+c_{2}V(r;W)+\frac{1}{\omega (W)}\int_{r}^{\infty
}[U(r;W)\otimes V(y;W)]F(y)dy+ \\
& +\frac{1}{\omega (W)}\int_{0}^{r}V(r;W)\otimes U(y;W)F(y)dy,
\end{align*}%
where $V(r;W)=V_{(1)}(r;W)$ and $\omega (W)$ are given by the respective
formulas (\ref{1.34}), (\ref{1.35}), and (\ref{1.36}). This representation
is a copy of representation (\ref{2.1.1}) with the change of $U_{(2)}\;$to $%
V_{(1)}$. It is correct because $V_{(1)}(r;W)$ with $\Im W>0$ decreases
exponentially as $r\rightarrow \infty $. The condition $\Psi \in \mathcal{L}%
^{2}(0,\infty )$, which is sufficient for $\Psi $ to belong to $D_{h}$
(because then automatically $\check{h}\Psi =W\Psi +F\in \mathcal{L}%
^{2}(0,\infty )$) implies that $c_{1}=$ $c_{2}=0$: otherwise, $\Psi $ is
non-square-integrable at infinity (if $c_1\neq0$) or at the origin (if $%
c_{2}\neq 0$) because $U(r;W)$ with $\Im W>0$ exponentially grows as $%
r\rightarrow \infty $ and $V(r;W)$ is non-square-integrable at the origin.
We thus obtain that the solution $\Psi \in D_{h\text{ }}$ of Eq. (\ref%
{2.2.0a}) with any $F\in D_{h\text{ }}$ is represented as 
\begin{align*}
&\Psi (r)=\frac{1}{\omega (W)}\int_{r}^{\infty }[U(r;W)\otimes V(y;W)]F(y)dy+
\\
&+\frac{1}{\omega (W)}\int_{0}^{r}[V(r;W)\otimes U(y;W)]F(y)dy,
\end{align*}
which is the required representation (\ref{2.2.0b}) with 
\begin{equation}
G(r,r^{\prime };W)=\left\{ 
\begin{array}{c}
\frac{1}{\omega (W)}V(r;W)\otimes U(r^{\prime };W),\;r>r^{\prime } \\ 
\frac{1}{\omega (W)}U(r;W)\otimes V(r^{\prime };W),\;r<r^{\prime }%
\end{array}%
\right. .  \label{2.2.1}
\end{equation}

This expression for Green's function allows evaluating the spectral function 
$\sigma (E)$ of the radial Hamiltonian $\hat{h}$ and writing the inversion
formulas in accordance with the instructions in Appendix~\ref{appA} (see
formulas (\ref{A1.1}) -- (\ref{A1.3c})).

\paragraph{Spectral function and inversion formulas}

According to (\ref{A1.3c}), (\ref{2.2.1}), and (\ref{1.35}) we obtain that%
\begin{align*}
& M(c;W)=\frac{1}{\omega (W)}U(c;W)\otimes V(c;W)= \\
& \,=\frac{1}{\omega (W)}U(c;W)\otimes U(c;W)+\frac{q}{2\gamma }%
U(c;W)\otimes U_{(2)}(c;W),
\end{align*}%
and because $U(c;E)=U_{(1)}(c;E)$ and $U_{(2)}(c;E)$ are real, formulas (\ref%
{A1.3a}) and (\ref{A1.3b}) then yield 
\begin{equation}
\frac{d\sigma (E)}{dE}=\frac{1}{\pi }\lim_{\varepsilon \rightarrow0} \Im%
\frac{1}{\omega (E+i\varepsilon )}  \label{3.1.4.1a}
\end{equation}%
for the spectral function $\sigma (E)$ of the radial Hamiltonian $\hat{h}$,
where $\lim $ in (\ref{3.1.4.1a}) is understood in a distribution theoretic
sense as well as $d\sigma (E)/dE$.

The spectral function is thus determined by the (generalized) function $%
\Im\omega ^{-1}(E)$,

\begin{equation*}
\omega^{-1}(E)=\lim_{\varepsilon\rightarrow0}\frac{1}{\omega(E+i\varepsilon)}%
.
\end{equation*}
At the points where the function $\omega(E)$, 
\begin{equation*}
\omega(E)=\lim_{\varepsilon\rightarrow0}\omega(E+i\varepsilon),
\end{equation*}
is different from zero, we have $\omega^{-1}(E)=1/\omega(E)$.

The explicit form of $\omega(W)$ (\ref{1.36}) shows that $\omega(E)$ exists
and is qualitatively different in the two energy regions $|E|\geq m$ and $%
|E|<m$. Therefore, we naturally distinguish these two energy regions in the
subsequent analysis.

We first consider the region $|E|\geq m$.

A direct verification shows that in this energy region, $\omega (E)$ is
continuous, different from zero, and takes complex values. It follows that
for\ $|E|\geq m$, the spectral function $\sigma (E)$ is absolutely
continuous and 
\begin{align}
& \frac{d\sigma (E)}{dE}=\frac{1}{\pi }\Im\frac{1}{\omega (E)}\equiv
Q^{2}(E),\;|E|\geq m,  \notag \\
& \omega (E)=\frac{2\gamma \Gamma (2\gamma )e^{\epsilon i\pi \gamma }\Gamma
(-\gamma +q|E|/ik)[(\varkappa +\gamma )\epsilon k+iq(E-m)](2k/m)^{-2\gamma }%
}{q\Gamma (-2\gamma )\Gamma (\gamma +q|E|/ik)[(\varkappa -\gamma )\epsilon
k+iq(E-m)]},\;  \label{3.1.4.2} \\
& \epsilon =E/|E|,\;k=\sqrt{E^{2}-m^{2}}.  \notag
\end{align}

We now consider the case of $|E|<m$.

In this energy region, we have 
\begin{align}
& \omega (E)=\frac{2\gamma \Gamma (2\gamma )\Gamma (-\gamma -qE/\tau
)[q(m-E)-(\varkappa +\gamma )\tau ](2\tau /m)^{-2\gamma }}{q\Gamma (-2\gamma
)\Gamma (\gamma -qE/\tau )[q(m-E)-(\varkappa -\gamma )\tau ]},
\label{3.1.4.2a} \\
& \tau =\sqrt{m^{2}-E^{2}},  \notag
\end{align}%
$\omega (E)$ is real, and $\lim_{\varepsilon \rightarrow 0}[1/\omega
(E+i\varepsilon )]\;$can be complex only at the points where $\omega (E)=0$.
Because $\Gamma (x)$ does not vanish for real $x$, $\omega (E)$ can vanish
only at the points satisfying one of the two conditions:

i) $q(m-E)-(\varkappa +\gamma )\tau =0$ or ii) $\gamma -qE/\tau =-n$, $%
\;n=0,1,...$, these are the points where $|\Gamma (\gamma -qE/\tau )|=\infty 
$.

The case i) yields$\;E=-\gamma m/\varkappa $ for$\;\zeta =1$, but at this
point, we also have $-\gamma -qE/\tau =0$, such that the product $\Gamma
(-\gamma -qE/\tau )[q(m-E)-(\varkappa +\gamma )\tau ]\neq 0$ and $\omega
(E)\neq 0$.

The case ii) yields $E=E_{n}=m/\sqrt{1+q^{2}/(n+\gamma )^{2}},\;n=0,1,...$,
but for$\;\zeta =1$at the point $\ E=E_{0}$\ , we also have $%
q(m-E)-(\varkappa -\gamma )\tau =0$, and consequently, $|\Gamma (\gamma
-qE/\tau )[q(m-E)-(\varkappa -\gamma )\tau ]|$ $<\infty $.

We thus obtain that $\omega (E)$ vanishes at the discrete points 
\begin{equation}
E=E_{n}=\frac{m}{\sqrt{1+\frac{q^{2}}{(n+\gamma )^{2}}}},\;n=\left\{ 
\begin{array}{l}
1,2,...,\;\zeta =1, \\ 
0,1,2,...,\;\zeta =-1,%
\end{array}%
\right.  \label{3.1.4.3}
\end{equation}%
which form the well-known discrete spectrum of bound states. We note that
the discrete spectrum accumulates at the point $E=m$, and its asymptotic
form as $n\rightarrow \infty $ is 
\begin{equation*}
\epsilon _{n}\equiv m-E_{n}=\frac{mq^{2}}{2n^{2}},
\end{equation*}
which is the well-known nonrelativistic formula for bound state energies.

In the vicinity of these points, we have 
\begin{equation*}
\frac{1}{\omega (E+i\varepsilon )}=-\frac{Q_{n}^{2}}{E-E_{n}+i\varepsilon }%
+O(1),\;Q_{n}^{2}=\lim_{E\rightarrow E_{n}}\frac{E_{n}-E}{\omega (E)}.
\end{equation*}%
It follows that for $|E|<m$, the spectral function $\sigma (E)$ is a jump
function with the jumps $Q_{n}^{2}$ located at the points $E=E_{n}$ (the
discrete energy eigenvalues(\ref{3.1.4.3})) and 
\begin{equation}
\frac{d\sigma (E)}{dE}=\sum_{n}Q_{n}^{2}\delta (E-E_{n}),\;n=\left\{ 
\begin{array}{l}
1,2,...,\;\zeta =1 \\ 
0,1,2,...,\;\zeta =-1%
\end{array}%
\right. ,\;|E|<m.  \label{3.1.4.3a}
\end{equation}

We finally obtain that the spectrum $\mathrm{Spec}\hat{h}$ of the operator $%
\hat{h}$\ is the union of the discrete spectrum $U_{n}\{E_{n}\}\subset
(-m,m) $ and the continuous spectrum containing the positive part $[m,\infty
)\;$and the negative part $(-\infty ,m]$, 
\begin{equation}
\mathrm{Spec}\hat{h}=(-\infty ,-m]\cup\left( \cup _{n}\{E_{n}\}\right) \cup
\lbrack m,\infty ).  \label{3.1.4.3b}
\end{equation}

We introduce a notation 
\begin{equation}
U_{\mathrm{norm}}(r;E)=\left\{ 
\begin{array}{l}
Q(E)U(r;E),\;|E|\geq m, \\ 
Q_{n}U(r;E_{n}),\;E=E_{n},\;|E|<m,%
\end{array}%
\right.  \label{3.1.4.3c}
\end{equation}

\begin{equation}
\varphi (E)=\left\{ 
\begin{array}{l}
Q(E)\Phi (E),\;|E|\geq m, \\ 
Q_{n}\Phi (E_{n}),\;E=E_{n},\;|E|<m.%
\end{array}%
\right.  \label{3.1.4.3d}
\end{equation}

The inversion formulas\ (\ref{A1.1}), (\ref{A1.2}) and Parseval equality (%
\ref{A1.2ab}) then become 
\begin{align}
& \varphi (E)=\int_{0}^{\infty }U_{\mathrm{norm}}(r;E)F(r)dr,\;E\in (-\infty
,-m]\cup(\cup _{n}\{E_{n}\})\cup \lbrack m,\infty ),  \label{3.1.4.4} \\
& F(r)=\int_{-\infty }^{-m}dEU_{\mathrm{norm}}(r;E)\varphi (E)+\sum_{n}U_{%
\mathrm{norm}}(r;E_{n})\varphi (E_{n})+\int_{m}^{\infty }dEU_{\mathrm{norm}%
}(r;E)\varphi (E),  \label{3.1.4.5} \\
& \int_{0}^{\infty }|F(r)|^{2}dr=\int_{-\infty }^{-m}|\varphi
(E)|^{2}dE+\sum_{n}|\varphi (E_{n})|^{2}+\int_{m}^{\infty }|\varphi
(E)|^{2}dE.  \label{3.1.4.5a} \\
& n=\left\{ 
\begin{array}{l}
1,2,...,\;\zeta =1 \\ 
0,1,2,...,\;\zeta =-1%
\end{array}%
\right. ,  \notag
\end{align}%
Inversion formulas and Parseval equality (\ref{3.1.4.4}) -- (\ref{3.1.4.5a})
are conventionally treated as the formulas for the generalized \ Fourier
expansion of doublets $F\in \mathcal{L}^{2}(0,\infty )$ with respect to the
complete orthonormalized set of the eigenfunctions $U_{\mathrm{norm}}(r;E)$
of s.a. radial Hamiltonian $\hat{h}$ (\ref{2.21}) associated with
s.a.differential expression $\check{h}$ (\ref{R.4}).

The obtained results for the energy spectrum and (generalized)
eigenfunctions coincide with the results obtained by the standard method
based on the physical arguments: the energy eigenstates must be locally
square integrable solutions of the differential equation $\check{h}F=EF$,
their moduli must be bounded at infinity, the eigenvalues $E$ corresponding
to the square-integrable bound eigenstates form the discrete energy
spectrum, and the non-square-integrable eigenstates corresponding to the
continuous energy spectrum must allow \ a ``normalization to $\delta $%
-function''.

As the first example, we apply these considerations to the energy region $%
|E|<m$. In this energy region,\ the solutions of the differential equation $%
\check{h}F=EF$ either exponentially grow or exponentially decrease (in\
addition,they\ can\ be\ non-square-integrable\ at\ the\ origin).\ Because
the required solutions must be locally square integrable, the energy
eigenstates must belong to $\mathcal{L}^{2}(0,\infty )$. It is convenient to
first find the solutions square integrable at infinity. They are given by (%
\ref{1.34})-(\ref{1.36}), 
\begin{equation*}
F(r)=cV_{(1)}(r;E)=c\left[ U_{(1)}(r;E)+\frac{q}{2\gamma }\omega
(E)U_{(2)}(r;E)\right] ,
\end{equation*}%
$c$ is a constant. These functions are square integrable at the origin and
therefore on the whole semiaxis only under the condition $\omega (E)=0$,
which reproduces the above results concerning the discrete spectrum and the
corresponding eigenfunctions.

As another example of an illustration of the standard method, we show that
by a direct calculation of the corresponding integrals, we can establish the
orthonormality relations for the eigenfunctions that are conventionally
represented in the physical literature as 
\begin{align}
& \int_{0}^{\infty }U_{\mathrm{norm}}(r;E_{n})U_{\mathrm{norm}}
(r;E_{n^{\prime }})dr=\delta _{nn^{\prime }},  \notag \\
& \int_{0}^{\infty }U_{\mathrm{norm}}(r;E_{n})U_{\mathrm{norm}}(r;E^{\prime
})dr=0,  \label{3.1.4.5b} \\
& \int_{0}^{\infty }U_{\mathrm{norm}}(r;E)U_{\mathrm{norm}}(r;E^{\prime
})dr=\delta (E-E^{\prime }),\;|E|,|E^{\prime }|\geq m.  \notag
\end{align}%
The method for calculating is presented in Appendix~\ref{appB} where it is
demonstrated by the example of the second noncritical charge region.
Unfortunately, we are unable to establish the the completeness relation for
the eigenfunctions that is conventionally written as 
\begin{align*}
& \int_{-\infty }^{-m}U_{\mathrm{norm}}(r;E)\otimes U_{\mathrm{norm}%
}(r^{^{\prime }};E)dE+\sum_{n}\sum U_{\mathrm{norm}}(r;E_{n})\otimes U_{%
\mathrm{norm}}(r^{^{\prime }};E_{n})+ \\
& +\int_{m}^{\infty }U_{\mathrm{norm}}(r;E)\otimes U_{\mathrm{norm}%
}(r^{^{\prime }};E)dE=\delta (r-r^{^{\prime }})I,
\end{align*}%
$I$ is the identity $2\times 2$ $\ $matrix, by a direct calculation of the
corresponding integrals, and we know no heuristic physical arguments in
support of the validity of this relation.

It now remains to consider the exceptional cases of $\gamma =n/2,$ $%
n=1,2,... $. As follows from Sec.~\ref{sec3}, in a neighborhood of each
point $\gamma=n/2$ \ and at the point itself, we can equivalently use the
doublets $U_{n(2)}$ and $V_{n(1)}$ (\ref{1.38b}) with the change of $%
\omega(W)$ to $\omega _{n}(W)$ ( see formulas (\ref{1.38a})\ and (\ref{1.38b}%
)) and obtain exactly the same conclusions about the guiding functional and
the same results for the Green's function, spectral function and
eigenfunctions as those for the case of $\gamma \neq n/2$. This evidently
follows from relations (\ref{1.39a}), (\ref{eq.27a}), and (\ref{1.39b}),
where, in particular, the term $\frac{q}{2\gamma }\Gamma (-2\gamma )A_{n}(W)$
in the r.h.s. in (\ref{1.39b}) is real for real $W=E.$ It also follows from
these formulas that Green's function and the spectral function are
continuous in $\gamma $ at each point $\gamma =n/2.$

\subsection{Second noncritical region}

\label{sec4.3}

This domain is characterized by the condition 
\begin{equation*}
q_{uj}<q<q_{cj}\Longleftrightarrow \mathrm{\;}0<\Upsilon _{+}=\gamma <\frac{1%
}{2}.
\end{equation*}

\subsubsection{Self-adjoint radial Hamiltonians}

\label{sec4.3.1}

As in the previous section, we first evaluate asymmetry form $\Delta _{\ast
}(F)$ (\ref{2.12}) evaluating the asymptotic behavior of the doublets $F\in
D_{h^{\ast }}$ at the origin with the use of representation \ (\ref{2.15})
-- (\ref{2.17}). In the case of $0<\gamma <1/2$ under consideration, the
both $U_{1}(r)\sim r^{\gamma }\;$and $U_{2}(r)\sim r^{-\gamma }$ are square
integrable at the origin and estimates (\ref{2.19}) hold true, such that for
any $F\in D_{h^{\ast }}$, we have 
\begin{align*}
& F(r)=c_{1}(mr)^{\gamma }u_{+}+c_{2}(mr)^{-\gamma
}u_{-}+O(r^{1/2}),\;r\rightarrow 0,\; \\
& u_{\pm }=\left( 
\begin{array}{c}
1 \\ 
\frac{\varkappa \pm \gamma }{q}%
\end{array}%
\right) ,
\end{align*}%
which in turn yields 
\begin{equation}
\Delta _{\ast }(F)=\frac{2\gamma }{q}(\overline{c_{2}}\,c_{1}-\overline{c_{1}%
}\,c_{2}).  \label{3.3.1.1}
\end{equation}%
The asymmetry form $\Delta _{\ast }(F)$ thus turns out to be a nontrivial
anti-Hermitian quadratic form in the asymptotic coefficients $c_{1}$ and $%
c_{2}$, which means that operator $\hat{h}^{\ast }\;$(\ref{2.6}) is not
symmetric and the problem of constructing nontrivial s.a. extensions of the
initial symmetric operator $\hat{h}^{(0)}$ (\ref{R.3}) arises.

In solving this problem, we follow a method in~\cite{22} that comprises two
steps.

1. Reducing the quadratic anti-Hermitian form $\,\Delta _{\ast }$ as\ a\
form\ in\ boundary values or asymptotic\ coefficients\ $c_{a}$, $a=1,2,...$,
to a canonical diagonal\ form by a linear transformation of the coefficients 
$c_{a}$ to coefficients $c_{+k\,}$,$\;k=1,...,m_{+}$, and $c_{-l}$, $%
l=1,...,m_{-}$, such that $\Delta _{\ast }$ becomes 
\begin{equation*}
\Delta _{\ast }=i\kappa (\sum_{1}^{m_{+}}\left| c_{+k\,}\right|
^{2}-\sum_{1}^{m_{-}}\left| c_{-l}\right| ^{2}),
\end{equation*}%
where $\kappa $ is some real coefficient.

2. Relating $c_{+k}$ and\ $c_{-l}$ by a unitary $\ m\times m$ matrix $U$, 
\begin{equation*}
c_{-l}=\sum_{1}^{m}U_{kl}c_{+k\text{ }},\;l=1,...,m,
\end{equation*}
if the inertia indices $m_{+}$ and $m_{-}$ of the form\footnote{%
The inertia indices coincide with the deficiency indices of an initial
symmetric operator.} are equal, $m_{+}=m_{-}=m$. Each such a relation with a
fixed $U$ convert the form $\Delta _{\ast}$ to zero and yields s.a.
(asymptotic) boundary conditions specifying a s.a. extension of the initial
symmetric operator, different $U$ define different s.a.extensions.
Conversely, any s.a. extension is specified by some $U$, and when $U$ runs
over the group $U(m)$, we obtain the whole $\ m^{2}$-parameter $U(m)$-family
of all possible s.a.extensions.

We apply this method to our case.

By a linear transformation 
\begin{equation*}
c_{1,2}\rightarrow c_{\pm }=c_{1}\pm ic_{2},
\end{equation*}%
the asymmetry form $\Delta _{\ast }$ is reduced to a canonical diagonal
form: 
\begin{equation*}
\Delta _{\ast }(F)=i\frac{\gamma }{q}(|c_{+}|^{2}-|c_{-}|^{2}).
\end{equation*}%
Its inertia indices are $(1,1)$, which in particular means that the
deficiency indices of $\hat{h}^{(0)}$ with $0<\gamma<1/2$ are $(1,1)$.

The relation 
\begin{equation}
c_{-}=e^{i\theta }c_{+},\;0\leq \theta \leq 2\pi ,\;0\sim 2\pi ,
\label{3.3.1.2}
\end{equation}%
\ with any fixed $\theta $ yields boundary conditions specifying a s.a.
extension $\hat{h}_{\theta \;}$of the operator $\hat{h}^{(0)}$. Different $%
\theta$'s are assigned different s.a. extensions, except equivalent cases of 
$\theta =0$ and $\theta =$ $2\pi $. When $\theta $ runs over a circle, we
obtain the whole one-parameter $U(1)$-family of all s.a. extensions of the
operator $\hat{h}^{(0)}$.

Relation (\ref{3.3.1.2}) is equivalent to the relation 
\begin{equation*}
c_{2}=\xi c_{1},\;-\infty \leq \xi =-\tan \frac{\theta }{2}\leq +\infty
,\;-\infty \sim +\infty ,
\end{equation*}%
the values $\xi =\pm \infty $ are equivalent and mean that $c_{1}=0$; we
will say that $\xi =\infty $ in these cases.

We let$\;\hat{h}_{\xi }$ redenote the corresponding s.a. operator, $\hat{h}%
_{\xi }\equiv \hat{h}_{\theta }$, and let $D_{\xi }$ denote its domain. The
final result in a more extended form is formulated as follows. In the second
noncritical region $0<\gamma <1/2$, we have a one-parameter $U(1)$-family $\{%
\hat{h}_{\xi }\}$ of s.a. operators associated with s.a. differential
expression $\check{h}$ (\ref{R.4}). They are specified by s.a. boundary
conditions and are given by 
\begin{equation}
\hat{h}_{\xi }:\left\{ 
\begin{array}{l}
D_{\xi }=\left\{ 
\begin{array}{l}
F(r):F(r)\text{\textrm{\ is absolutely continuous in} }(0,\infty ),\;F,%
\check{h}F\subset \mathcal{L}^{2}(0,\infty ), \\ 
F(r)=c[(mr)^{\gamma }u_{+}+\xi (mr)^{-\gamma
}u_{-}]+O(r^{1/2}),\;r\rightarrow 0,\;-\infty <\xi <+\infty , \\ 
F(r)=c(mr)^{-\gamma }u_{-}+O(r^{1/2}),\;r\rightarrow 0,\;\xi =\infty ,%
\end{array}%
\right\} \\ 
\hat{h}_{\xi }F=\check{h}F,%
\end{array}%
\right.  \label{3.3.1.3}
\end{equation}%
where $c$ is an arbitrary complex number.

In other words, the only s.a. differential expression$\ \check{h}$ does not
uniquely define a s.a. operator in the charge region $0<\gamma <1/2$, and an
additional specification of the domain in terms of s.a asymptotic boundary
conditions involving one real parameter $\xi $ is required.

\subsubsection{Spectral analysis}

\label{sec4.3.2}

The spectral analysis in this charge region is quite similar to the analysis
performed in Subsec.~\ref{sec4.2} related to the first noncritical region%
\footnote{%
A simplifying thing is that the particular cases of $\gamma =0$ and $\gamma
=1/2$ are excluded.}. We therefore only point out necessary modifications
and formulate the final results.

In the case of $\xi =0$, the the corresponding analysis is identical to that
in the previous Subsec.~\ref{sec4.2} and the results obtained there are
directly extended to the region $0<$ $\gamma <1/2$ and are given by the same
formulas.

Until said otherwise, we assume that $0<|\xi |<\infty $, $\xi $ is arbitrary
, but fixed. The case of $\xi =\infty $ is considered separately below.

For the doublet $U(r;W)$\ defining guiding functional (\ref{A1.0}) , we
choose the doublet 
\begin{equation}
U_{\xi }(r;W)=U_{(1)}(r;W)+\xi U_{(2)}(r;W)  \label{3.3.2.1}
\end{equation}%
satisfying the condition

\begin{equation*}
U_{\xi }(r;W)=(mr)^{\gamma }u_{+}+\xi (mr)^{-\gamma }u_{-}+O(r^{-\gamma
+1}),\;r\rightarrow 0,
\end{equation*}%
where $U_{(1)}$ and $U_{(2)}$ are given by formulas (\ref{1.27}) -- (\ref%
{1.33}). As before, $U_{\xi }(r;W)$ is real-entire in $W$. The corresponding
guiding functional is denoted by $\Phi _{\xi }(F,W)$.

For $\mathcal{D}$, we choose the set $\mathcal{D}_{\xi }$ of doublets
belonging to $D_{\xi}$ and having a compact support.

The guiding functional $\Phi _{\xi }$ with the chosen $U_{\xi }$ and $%
\mathcal{D}_{\xi \;}$is simple. Indeed, the properties 1) and 3) are
evident; as to the property 2), the solution $\Psi$ of the inhomogeneous
equation $(\check{h}-E_{0})\Psi =F_{0}$, $F_{0}\in\mathcal{D}_{\xi}$, with
the property $\Psi\in\mathcal{D}_{\xi}$ is given by a copy of (\ref{2.1.2}),
where the solutions $U=U_{(1)}$ and $U_{(2)}$ of the homogeneous equation
are replaced by the respective solutions $U_{\xi }$ and $U_{(1)}$ with the
Wronskian $\mathrm{Wr}(U_{\xi },U_{(1)})=2\gamma\xi/q$, 
\begin{align*}
& \Psi (r)=-\frac{q}{2\gamma \xi }\int_{r}^{\infty }[U_{\xi
}(r;E_{0})\otimes U_{(1)}(y;E_{0})]F_{0}(y)dy+ \\
& -\frac{q}{2\gamma \xi }\int_{0}^{r}U_{(1)}(r;E_{0})\otimes U_{\xi
}(y;E_{0})]F_{0}(y)dy.
\end{align*}

Green's function $G_{\xi }(r,r^{\prime };W)$ , $\Im W>0$, of the operator $%
\hat{h}_{\xi }$ is defined as the kernel of the integral representation for
any $\Psi \in D_{\xi }$ in terms of the doublet $F=(\hat{h}_{\xi }-W)\Psi\in%
\mathcal{L}^{2}(0,\infty)$: 
\begin{equation*}
\Psi (r)=\int_{0}^{\infty }G_{\xi }(r,r^{\prime };W)F(r^{\prime })dr^{\prime
}.
\end{equation*}
This representation is a copy of formula (\ref{2.2.0b}). The natural
difference is the change of $U=U_{(1)}$ to $U$ $=U_{\xi }$ because the
condition $\Psi \in D_{\xi }$ implies that $\Psi $ satisfies s.a. asymptotic
boundary conditions (\ref{3.3.1.3}). The final result is 
\begin{equation*}
G_{\xi }(r,r^{\prime };W)=\left\{ 
\begin{array}{c}
\frac{1}{\omega _{\xi }(W)}V(r;W)\otimes U_{\xi }(r^{\prime
};W),\;r>r^{\prime }, \\ 
\frac{1}{\omega _{\xi }(W)}U_{\xi }(r;W)\otimes V(r^{\prime
};W),\;r<r^{\prime },%
\end{array}%
\right.
\end{equation*}%
where 
\begin{align}  \label{3.3.2.2}
\begin{aligned} &V(r;W)=V_{(1)}(r;W)=U_\xi(r;W)+\frac{q}{2\gamma}
\omega_\xi(W)U_{(2)}(r;W), \\
&\omega_\xi(W)=-\operatorname{Wr}(U,V)=\omega(W)-\frac{2\gamma\xi}{q},
\end{aligned}
\end{align}
$V_{(1)}(r;W)$ and $\omega (W)$ are given by (\ref{1.34}) -- (\ref{1.36}),
which is a copy of representation (\ref{2.2.1})\ with the natural change of $%
U=U_{(1)}\ $and $\omega $ to the respective $U$ $=U_{\xi }$ and $\omega
_{\xi }$.

It follows, see (\ref{A1.3a}) -- (\ref{A1.3c}), that 
\begin{align*}
& M(c;W)=\frac{1}{\omega _{\xi }(W)}U_{\xi }(c;W)\otimes V(c;W)= \\
& =\frac{1}{\omega _{\xi }(W)}U_{\xi }(c;W)\otimes U_{\xi }(c;W)+\frac{q}{%
2\gamma }U_{(2)}(c;W)\otimes U_{\xi }(c;W),\;
\end{align*}%
and, because \ the both $U_{\xi }(c;E)$ and $U_{(2)}(c;E)$ are real, that
the spectral function $\sigma _{\xi }(E)$ of the radial Hamiltonian $\hat{h}%
_{\xi }$, $0<|\xi |<\infty $, is given by 
\begin{equation*}
\frac{d\sigma _{\xi }(E)}{dE}=\frac{1}{\pi }\lim_{\varepsilon \rightarrow
0}\Im\frac{1}{\omega _{\xi }(E+i\varepsilon )},
\end{equation*}%
a copy of expression (\ref{3.1.4.1a})\ with the change of $\omega $ to $%
\omega _{\xi }$. The spectral function is determined by the (generalized)
function $\Im\omega _{\xi }^{-1}(E)$,

\begin{equation*}
\omega _{\xi }^{-1}(E)=\lim_{\varepsilon \rightarrow 0}\frac{1}{\omega _{\xi
}(E+i\varepsilon )}=\lim_{\varepsilon \rightarrow 0}\frac{1}{\omega
(E+i\varepsilon )-2\gamma \xi /q}.
\end{equation*}%
At the points where the function 
\begin{equation*}
\omega _{\xi }(E)=\lim_{\varepsilon \rightarrow 0}\omega _{\xi
}(E+i\varepsilon )=\lim_{\varepsilon \rightarrow 0}[\omega (E+i\varepsilon
)-2\gamma \xi /q]
\end{equation*}%
is different from zero, we have $\omega _{\xi }^{-1}(E)=1/\omega _{\xi }(E)$%
. Because $\omega _{\xi }(E)$ differs from $\omega (E)$ by a real constant $%
-2\gamma \xi /q$, the two energy regions $|E|\geq m\,$and $|E|<m\,$are
naturally distinguished as before, and the corresponding analysis in each
region is similar to the analysis in the previous subsection. In the energy
region $|E|\geq m$, the function $\omega _{\xi }(E)$ is continuous,
different from zero, and complex as well as $\omega (E)$ (\ref{3.1.4.2}).
The spectral function $\sigma _{\xi }(E)$ for $|E|\geq m$ is therefore
absolutely continuous, and 
\begin{equation*}
\frac{d\sigma _{\xi }(E)}{dE}=\frac{1}{\pi }\Im\frac{1}{\omega (E)-\frac{%
2\gamma \xi }{q}}\equiv Q_{\xi }^{2}(E),
\end{equation*}%
which is an analogue of (\ref{3.1.4.2}) with the change of $\omega (E)$ and $%
Q(E)$ to the respective $\omega _{\xi }(E)$ and $Q_{\xi }(E)$.

In the energy region $\ |E|<m$, function $\omega (E)$ (\ref{3.1.4.2a})$\ $is
real, and therefore, the function $\omega _{\xi }(E)$ is also real. As in
the case of the first noncritical charge region, it follows that for $|E|<m$%
, the spectral function $\sigma _{\xi }(E)$ is a jump function with the
jumps $Q_{\xi ,n}^{2}$ located at the points $E=E_{\xi ,n}$, the discrete
energy eigenvalues,\ where $\omega _{\xi }(E_{\xi ,n})=0$, and 
\begin{equation*}
Q_{\xi ,n}^{2}=\lim_{E\rightarrow E_{\xi ,n}}\frac{E_{\xi ,n}-E}{\omega
_{\xi }(E)}.
\end{equation*}%
As a result, we obtain that 
\begin{equation*}
\frac{d\sigma _{\xi }(E)}{dE}=\sum_{n}Q_{\xi ,n}^{2}\delta (E-E_{\xi
,n}),\;|E|<m,
\end{equation*}%
which is an analogue of (\ref{3.1.4.3a}) with the change of $E_{n}$ and $%
Q_{n}$ to the respective $E_{\xi ,n}$ and $Q_{\xi ,n}$.

Unfortunately, we are unable to find an explicit formula for the discrete
energy eigenvalues $E_{\xi ,n}$ with $\xi \neq 0$, we only note that, as in
the first noncritical charge region, there are infinitely many of such
levels accumulating at the point $E=m$, and their asymptotic form as $%
n\rightarrow \infty $ is given by the previous nonrelativistic expression
independent of $\xi$: 
\begin{equation*}
\epsilon _{\xi ,n}\equiv m-E_{\xi,n}= \frac{mq^{2}}{2n^{2}}.
\end{equation*}

The lower bound state energy essentially depends on $\xi$, and there exists
a value of $\xi $ for which the lower bound state energy coincides with the
boundary $E=-m$ of the lower (positron) continuous spectrum.

The whole spectrum \text{Spec}$\hat{h}_{\xi }$ of the radial Hamiltonian $%
\hat{h}_{\xi }$ is given by a copy of (\ref{3.1.4.3b})\ with the change of $%
E_{n}$ to $E_{\xi ,n}$.

The inversion formulas and the Parseval equality 
\begin{align*}
& \varphi _{\xi }(E)=\int_{0}^{\infty }U_{\xi ,\mathrm{norm}%
}(r;E)F(r)dr,\;E\in (-\infty ,-m]\cup(\cup _{n}\{E_{\xi ,n}\})\cup \lbrack
m,\infty ), \\
& F(r)=\int_{-\infty }^{-m}dEU_{\xi ,\mathrm{norm}}(r;E)\varphi _{\xi
}(E)+\sum_{n}U_{\xi ,\mathrm{norm}}(r;E_{\xi ,n})\varphi _{\xi }(E_{\xi ,n})+
\\
& +\int_{m}^{\infty }dEU_{\xi ,\mathrm{norm}}(r;E)\varphi _{\xi }(E), \\
& \int_{0}^{\infty }|F(r)|^{2}dr=\int_{-\infty }^{-m}|\varphi _{\xi
}(E)|^{2}dE+\sum_{n}|\varphi _{\xi }(E_{\xi ,n})|^{2}+\int_{m}^{\infty
}|\varphi _{\xi }(E)|^{2}dE,
\end{align*}%
are written in terms of the normalized eigenfunctions $U_{\xi ,\mathrm{norm}%
}(r;E)$ and\ the Fourier coefficients $\varphi _{\xi }(E)$ that are defined
by copies of formulas (\ref{3.1.4.3c}) and (\ref{3.1.4.3d}) with the
addition of the subscript $\xi $ to all the symbols $Q$, $U$, $Q_{n}$, $%
E_{n} $, and $\Phi $. These relations are copies of formulas (\ref{3.1.4.4}%
)-(\ref{3.1.4.5a}).

As before, the energy spectrum and (generalized) eigenfunctions of the
radial Hamiltonian $\hat{h}_{\xi }$ can be obtained by the standard method
using the physical arguments.

As an example, we consider the energy region $|E|<m$ where the solutions $F$
of the differential equation $\check{h}F=EF$ either exponentially grow or
exponentially decrease as $r\rightarrow \infty $, all solutions being square
integrable at the origin. Only exponentially decreasing solutions

\begin{align*}
& F(r)=cV_{(1)}(r;E)=c[U_{(1)}(r;E)+\frac{q}{2\gamma }\omega
(E)U_{(2)}(r;E)]= \\
& \,=c[U_{\xi }(r;E)+\frac{q}{2\gamma }\omega _{\xi }(E)U_{(2)}(r;E)],
\end{align*}%
$c$ is a constant, are proper. It is remarkable that such solutions are
square-integrable on the whole semiaxis for any energy values $E\in (-m,m)$.
But they satisfy s.a. asymptotic boundary conditions (\ref{3.3.1.3})\ only
if $\omega _{\xi }(E)=0$, which reproduces the results for the eigenvalues
and eigenfunctions of the discrete spectrum of the operator $\hat{h}_{\xi }$%
. We note that the s.a. boundary conditions have a physical meaning of the
condition that probability flux density vanish at the boundary, the origin
in our case . We did not refer to this requirement in the first noncritical
charge region because it is automatically satisfied in this region.

We can also establish the orthonormality relations for the eigenfunctions $%
U_{\xi ,\mathrm{norm}}(r;E)$, which are copies of relations (\ref{3.1.4.5b}%
), by a direct calculation of the corresponding integrals. This calculation
illustrating the general method applicable to all the charge regions is
presented in Appendix~\ref{appB}.

As before, we are unable to establish the completeness relation for the
eigenfunctions by a direct calculation or based on heuristic physical
arguments.

We now touch briefly on the case of $\xi =\infty $ where the s.a. asymptotic
boundary conditions for any$\;F\in D_{\infty }$are 
\begin{equation*}
F(r)=c(mr)^{-\gamma }u_{-}+O(r^{1/2}),\;r\rightarrow 0.
\end{equation*}%
For the doublets $U$ and $V$, we choose the respective 
\begin{align*}
& U(r;W)=U_{\infty }(r;W)=U_{(2)}(r;W), \\
& U_{\infty }(r;W)=(mr)^{-\gamma }u_{-}+O(r^{1-\gamma }),\;r\rightarrow 0,
\end{align*}%
and 
\begin{align*}
V(r;W)& =\frac{2\gamma }{q\omega (W)}V_{(1)}(r;W)=U_{(2)}(r;W)-\frac{q}{%
2\gamma }\omega _{\infty }(W)U_{(1)}(r;W), \\
\omega _{\infty }(W)& =-\mathrm{Wr}(U,V)=-\frac{4\gamma ^{2}}{q^{2}\omega (W)%
}.
\end{align*}

Performing calculations completely similarly to those in the case of $|\xi
|<\infty $, we find the spectral function $\sigma _{\infty }(E)$, 
\begin{equation*}
\frac{d\sigma _{\infty }(E)}{dE}=\frac{1}{\pi }\lim_{\varepsilon \rightarrow
0}\Im\frac{1}{\omega _{\infty }(E+i\varepsilon )}=-\frac{1}{\pi }\frac{q^{2}%
}{4\gamma ^{2}}\lim_{\varepsilon \rightarrow 0}\Im\omega (E+i\varepsilon ).
\end{equation*}

All other results concerning the structure of the spectrum and inversion
formulas are also completely similar to the results in the case of $|\xi
|<\infty $ . In particular, the bound state spectrum is determined by the
poles of the function $\omega (E)$ in the energy region $|E|<m$; it can be
evaluated explicitly.

In conclusion, we note that the spectrum and the normalized eigenfunctions $%
U_{\xi ,\mathrm{norm}}(r;E)$ are continuous in $\xi $ , the points $\xi =0$
and $\xi =\infty $ included.

\subsection{Critical charges}

\label{sec4.4}

This region is defined by the charge values 
\begin{equation*}
q=q_{cj}=|\varkappa |=j+1/2\mathrm{\ \Longleftrightarrow }\;\Upsilon
_{+}=\gamma =0.
\end{equation*}%
The charge values $q=q_{cj}\;$stand out because for $q>q_{cj}$,\ the
standard formula (\ref{3.1.4.3}) for the bound state spectrum ceased to be
true yielding complex energy values. But we will see that from the
mathematical standpoint, no extraordinary happens with the system for the
charge values $q\geq q_{cj}$ , at least, in comparison with the previous
case of $q_{uj}<q<q_{cj}$.

\subsubsection{Self-adjoint radial Hamiltonians}

\label{sec4.4.1}

In constructing the s.a. Hamiltonians with critical charges, we literally
follow the method presented in the previous Subsec.~\ref{sec4.3} where the
second noncritical charge region was considered and therefore restrict
ourselves to only key remarks.

It follows from representation (\ref{2.15}) -- (\ref{2.17}) that the
asymptotic behavior of the doublets $F\in D_{\ast }$ in the case of $%
q=q_{cj} $ is given by 
\begin{equation*}
F(r)=c_{1}u_{+}+c_{2}u_{-}^{(0)}(r)+O(r^{1/2}\ln r),\;r\rightarrow
0,\;\forall F\in D_{h^{\ast }},
\end{equation*}%
which yields the expression 
\begin{equation*}
\Delta _{\ast }(F)=\frac{1}{q_{cj}}(\overline{c_{1}}c_{2}-\overline{c_{2}}%
c_{1})
\end{equation*}%
for $\Delta _{\ast }(F)$, this expression is completely similar to the
previous case, see (\ref{3.3.1.1}) with the changes $2\gamma /q\rightarrow
1/q_{cj}$ and $c_{1}\leftrightarrows c_{2}$.

Therefore, in the case of critical charges, i.e., for $\gamma =0$, we also
have the one-parameter $U(1)$-family $\{\hat{h}_{(0)\xi }\}$, $-\infty \leq
\xi \leq +\infty $,\ of s.a. operators associated with s.a.differential
expression $\check{h}$ (\ref{R.4}) and specified by s.a. asymptotic boundary
conditions, 
\begin{equation}
\hat{h}_{(0)\xi }:\left\{ 
\begin{array}{l}
D_{\xi }^{(0)}=\left\{ 
\begin{array}{l}
F(r):F(r)\text{\textrm{\ is absolutely continuous in} }(0,\infty ),\;F,%
\check{h}F\subset \mathcal{L}^{2}(0,\infty ), \\ 
F(r)=c(u_{-}^{(0)}(r)+\xi u_{+})+O(r^{1/2}\ln r),\;r\rightarrow 0,\;-\infty
<\xi <+\infty , \\ 
F(r)=cu_{+}+O(r^{1/2}\ln r),\;r\rightarrow 0,\;\xi =\infty ,%
\end{array}%
\right\} \\ 
\hat{h}_{(0)\xi }F=\check{h}F,%
\end{array}%
\right.  \label{3.4.1.1}
\end{equation}%
$D_{\xi }^{(0)}$ denotes the domain of the operator $\hat{h}_{(0)\xi }$, and 
$\xi =\infty $\ corresponds to the equivalent cases of $\xi=+\infty $ and $%
\xi =-\infty $.

\subsubsection{Spectral analysis}

\label{sec4.4.2}

The spectral analysis follows the standard way presented in the previous
subsections, and therefore, we only cite the final results.

We first consider the case of $\xi \neq \infty $. For the doublet $U(r;W)$
defining the guiding functional (\ref{A1.0}), we choose the doublet 
\begin{equation*}
U_{\xi }^{(0)}(r;W)=U_{(2)}^{(0)}(r;W)+\xi U_{(1)}(r;W)
\end{equation*}%
with the asymptotic behavior%
\begin{equation*}
U_{\xi }^{(0)}(r;W)=u_{-}^{(0)}(r)+\xi u_{+}+O(r\ln r),\;r\rightarrow 0,
\end{equation*}%
where the doublets $U_{(1)}$ and $U_{(2)}^{(0)}$ are given by formulas (\ref%
{1.40}) -- (\ref{1.42a}); $U_{\xi }^{(0)}(r;W)$ is real-entire in $W$.

For $\mathcal{D}$, we choose the set $\mathcal{D}_{\xi }^{(0)}$ of doublets
belonging to $D_{\xi }^{(0)}$ and having a compact support.

The guiding functional $\Phi _{\xi }^{(0)}$ with these $U_{\xi }^{(0)}$ and $%
\mathcal{D}_{\xi }^{(0)}$ is simple. In particular, the solution $\Psi%
\mathcal{D}_{\xi }^{(0)}$ of the inhomogeneous equation\ $(\check{h}%
-E_{0})\Psi =F_{0}$, $F_{0}\in \mathcal{D}_{\xi }^{(0)}$, is given by a copy
of (\ref{2.1.2}) with the change of $U_{(1)},U_{(2)}$, and $\frac{q}{2\gamma 
}=-(\mathrm{Wr}(U_{(1)},U_{(2)})^{-1}$ to the respective $U_{\xi }^{(0)}$, $%
U_{(1)}$, and $-q_{cj}=-(\mathrm{Wr}(U_{\xi }^{(0)},U_{(1)})^{-1}$.

Green's function $G_{\xi }^{(0)}(r,r^{\prime };W)$, $\Im W>0$, of the
Hamiltonian $\hat{h}_{(0)\xi }$ is defined by a copy of (\ref{2.2.1}) with
the change of $U=U_{(1)}$, $V=V_{(1)}$, and $\omega =$ $-\mathrm{Wr}%
(U_{(1)},V_{(1)})$ to the respective $U=U_{\xi }^{(0)},\;V=V_{(1)}^{(0)}\,$,
and $\omega _{\xi }^{(0)}=$ $-\mathrm{Wr}(U_{\xi }^{(0)},V_{(1)}^{(0)})$, 
\begin{equation*}
G_{\xi }^{(0)}(r,r^{\prime };W)=\left\{ 
\begin{array}{c}
\frac{1}{\omega _{\xi }^{(0)}(W)}V_{(1)}^{(0)}(r;W)\otimes U_{\xi
}^{(0)}(r^{\prime };W),\;r>r^{\prime }, \\ 
\frac{1}{\omega _{\xi }^{(0)}(W)}U_{\xi }^{(0)}(r;W)\otimes
V_{(1)}^{(0)}(r^{\prime };W),\;r<r^{\prime },%
\end{array}%
\right.
\end{equation*}%
see formulas (\ref{1.43})\ and (\ref{1.44}); $V_{(1)}^{(0)}$ is conveniently
represented as\ 
\begin{equation}
V_{(1)}^{(0)}(r;W)=U_{\xi }^{(0)}(r;W)+q_{cj}\omega _{\xi
}^{(0)}(W)U_{(1)}(r;W),  \label{3.4.1.2}
\end{equation}%
where 
\begin{equation*}
\omega _{\xi }^{(0)}(W)=-\mathrm{Wr}(U_{\xi }^{(0)},V_{(1)}^{(0)})=\omega
^{(0)}(W)-\frac{\xi }{q_{cj}},
\end{equation*}%
$\omega ^{(0)}(W)$ is given in (\ref{1.44}).

The spectral function $\sigma _{\xi }^{(0)}(E)$ of the radial Hamiltonian $%
\hat{h}_{(0)\xi }\,$is defined by 
\begin{equation*}
\frac{d\sigma _{\xi }^{(0)}(E)}{dE}=\frac{1}{\pi }\lim_{\varepsilon
\rightarrow 0}\Im\frac{1}{\omega _{\xi }^{(0)}(E+i\varepsilon )}
\end{equation*}%
and is determined by the (generalized) function $\Im\omega _{\xi
}^{(0)\,-1}(E)$, 
\begin{equation*}
\omega _{\xi }^{(0)\,-1}(E)=\lim_{\varepsilon \rightarrow 0}\frac{1}{\omega
_{\xi }^{(0)}(E+i\varepsilon )}=\lim_{\varepsilon \rightarrow 0}\frac{1}{%
\omega ^{(0)}(E+i\varepsilon )-\xi /q_{cj}}.
\end{equation*}%
At the points where the function 
\begin{equation*}
\omega _{\xi }^{(0)}(E)=\lim_{\varepsilon \rightarrow 0}\omega _{\xi
}^{(0)}(E+i\varepsilon )=\omega ^{(0)}(E)-\xi /q_{cj}
\end{equation*}
\ with%
\begin{equation*}
\omega ^{(0)}(E)=\lim_{\varepsilon \rightarrow 0}\omega
^{(0)}(E+i\varepsilon ),
\end{equation*}%
is different from zero, we have $\omega _{\xi }^{(0)\,-1}(E)=1/\omega _{\xi
}^{(0)}(E)$.

The two energy regions $|E|\geq m\,$and $|E|<m$ are naturally distinguished
as before.

In the region $|E|\geq m$, the function $\omega ^{(0)}(E)$ is given by 
\begin{align*}
& \omega ^{(0)}(E)=\frac{1}{q_{cj}}\left\{ \ln [2e^{-i\epsilon \pi
/2}k/m]+\psi (-iq_{cj}|E|/k)+\frac{i\epsilon k-\zeta m}{2q_{cj}E}-2\psi
(1)+\zeta /(2q_{cj})\right\} , \\
& \epsilon =E/|E|,\;k=\sqrt{E^{2}-m^{2}}.
\end{align*}%
It is continuous, different from zero, and complex, therefore, the spectral
function $\sigma _{\xi }^{(0)}(E)$ for $|E|\geq m$ is absolutely continuous,
and 
\begin{equation*}
\frac{d\sigma _{\xi }^{(0)}(E)}{dE}=\frac{1}{\pi }\Im\frac{1}{\omega
^{(0)}(E)-\frac{\xi }{q_{cj}}}\equiv \lbrack Q_{\xi }^{(0)}(E)]^{2}.
\end{equation*}

In the region$\ |E|<m$, the function $\omega ^{(0)}(E)$ is given by 
\begin{align*}
& \omega ^{(0)}(E)=\frac{1}{q_{cj}}\left\{ \ln (2\tau /m)+\psi
(-q_{cj}E/\tau )-\frac{\tau +\zeta m}{2q_{cj}E}-2\psi (1)+\zeta
/(2q_{cj})\right\} , \\
& \tau =\sqrt{m^{2}-E},
\end{align*}%
it$\ $is real, and therefore, the function $\omega _{\xi }^{(0)}(E)$ is also
real. As in the previous cases, the spectral function $\sigma _{\xi
}^{(0)}(E)$ for$\ |E|<m$ is a jump function with the jumps 
\begin{equation*}
\lbrack Q_{\xi ,n}^{(0)}]^{2}=\lim_{E\rightarrow E_{\xi ,n}^{(0)}}\frac{%
E_{\xi ,n}^{(0)}-E}{\omega _{\xi }^{(0)}(E)}
\end{equation*}%
located at the discrete energy eigenvalues $E_{\xi ,n}^{(0)}$ where $\omega
_{\xi }^{(0)}(E_{\xi ,n}^{(0)})=0$, such that 
\begin{equation*}
\frac{d\sigma _{\xi }^{(0)}(E)}{dE}=\sum_{n}[Q_{\xi ,n}^{(0)}]^{2}\delta
(E-E_{\xi ,n}^{(0)}).
\end{equation*}

As in the previous case, we are unable to find an explicit formula for $%
E_{\xi ,n}^{(0)}$ (except the case of $\xi =\infty $, see below), we only
note that there exists an infinite number of of the discrete levels which
accumulate at the point $E=m$, and their asymptotic behavior as $%
n\rightarrow \infty $ is described by the same nonrelativistic formula: 
\begin{equation*}
\epsilon _{\xi ,n}^{(0)}\equiv m-E_{\xi,n}^{(0)}=\frac{mq^{2}}{2n^{2}}.
\end{equation*}

The lower bound state energy essentially depends on $\xi $, and there exists
a value of $\xi $ for which the lower bound state energy coincides with the
boundary $E=-m$ of the lower (positron) continuous spectrum.

All other results concerning the inversion formulas and Parseval equality
are written in terms of the normalized (generalized) eigenfunctions $U_{\xi ,%
\mathrm{norm}}^{(0)}(r;E)\;$and\ the Fourier coefficients $\varphi _{\xi
}^{(0)}(E)$ as copies of relations (\ref{3.1.4.3c})-(\ref{3.1.4.5a}) with
the addition of the subscript $\xi $ and superscript $(0)$ to all the
symbols $Q$, $U$, $Q_{n}$, $E_{n}$, and $\Phi $.

As before, the energy spectrum and eigenfunctions of the radial Hamiltonian $%
\hat{h}_{(0)\xi }$ can be obtained by the standard method using the physical
arguments. As an example, we again consider the energy region $|E|<m$ where
the solutions $F$ of the differential equation $\check{h}F=EF$ either
exponentially grow or exponentially decrease and are square integrable at
the origin. The exponentially decreasing solutions are described by the
doublets 
\begin{equation*}
F(r)=cV_{(1)}^{(0)}(r;E)=c[U_{\xi }^{(0)}(r;E)+q_{cj}\omega _{\xi
}^{(0)}(E)U_{(1)}(r;E)],
\end{equation*}%
where $c$ is a constant. They are square integrable, $F\in \mathcal{L}%
^{2}(0,\infty )$, for any $E\;$\ in the interval $E\in (-m,m)$, but satisfy
s.a. asymptotic boundary conditions (\ref{3.4.1.1})\ only if $\omega_{\xi
}^{(0)}(E)=0$, which reproduces the results for the eigenvalues and
eigenfunctions of the discrete spectrum. We can also verify the
orthonormality relations for the eigenfunctions $U_{\xi,\mathrm{norm}%
}^{(0)}(r;E)$ that are analogues of relations (\ref{3.1.4.5b}) by a direct
calculation of the corresponding integrals with the method described in
Appendix~\ref{appB}.

We touch briefly on the case of $\xi =\infty $\ where the s.a asymptotic
boundary conditions for any $\forall F\in D_{\infty }$ are 
\begin{equation*}
F(r)=cu_{+}+O(r^{1/2}\ln r),\;r\rightarrow 0.
\end{equation*}%
For the doublets $U$ and $V$, we choose the respective 
\begin{align*}
& U(r;W)=U_{\infty }^{(0)}(r;W)=U_{(1)}(r;W), \\
& U_{\infty }^{(0)}(r;W)=cu_{+}+O(r),\;r\rightarrow 0,
\end{align*}%
and 
\begin{align*}
V(r;W)& =\frac{1}{q_{cj}\omega ^{(0)}(W)}%
V_{(1)}^{(0)}(r;W)=U_{(2)}^{(0)}(r;W)-q_{cj}\omega _{\infty
}^{(0)}(W)U_{(1)}(r;W), \\
\omega _{\infty }^{(0)}(W)& =-\mathrm{Wr}(U_{(1)},V)=-\frac{1}{%
q_{cj}^{2}\omega ^{(0)}(W)}.
\end{align*}%
Proceeding completely similarly to the case of $|\xi |<\infty $, we find the
spectral function $\sigma _{\infty }^{(0)}(E)$: 
\begin{equation*}
\frac{d\sigma _{\infty }^{(0)}(E)}{dE}=\frac{1}{\pi }\lim_{\varepsilon
\rightarrow 0}\Im\frac{1}{\omega _{\infty }^{(0)}(E+i\varepsilon )}=-\frac{1%
}{\pi }q_{cj}^{2}\lim_{\varepsilon \rightarrow 0}\Im\omega
^{(0)}(E+i\varepsilon ).
\end{equation*}

The structure of the spectrum, the inversion formulas, and the
orthonormality relations are also completely similar to the corresponding
results in the case of $|\xi |<\infty $ . In particular, the bound state
spectrum is determined by the poles of the function $\omega^{(0)}(E)$ in the
interval $|E|<m$. It can be evaluated explicitly and is given by formula (%
\ref{3.1.4.3}) with $\gamma =0$, the energy of the lower level with $\zeta
=-1$ is equal to zero, $E_{\infty ,0}^{(0)}=0$.

In conclusion, we note that the spectrum and the normalized eigenfunctions $%
U_{\xi ,\mathrm{norm}}^{(0)}(r;E)$ are continuous in $\xi$, the point $%
\xi=\infty $ included.

\subsection{Overcritical charges}

\label{sec4.5}

This region is defined by the charge values 
\begin{equation*}
q>q_{cj}=|\varkappa |=j+1/2\text{\ }\Longleftrightarrow \text{ }\Upsilon
_{+}=i\sigma ,\;\sigma =\sqrt{q^{2}-\varkappa ^{2}}>0.
\end{equation*}

In constructing s.a. Hamiltonians and analyzing their spectral properties in
this charge region, we canonically follow the methods used in the previous
cases and therefore only cite the main results.

\subsubsection{Self-adjoint\ radial Hamiltonians}

\label{sec4.5.1}

According to representation (\ref{2.15}) -- (\ref{2.17}), the asymptotic
behavior of the doublets $F\in D_{\ast}$ in the case of $q>q_{cj}$ is given
by 
\begin{align*}
&
F(r)=c_{1}(mr)^{i\sigma}u_{+}+c_{2}(mr)^{-i\sigma}u_{-}+O(r^{1/2}),\;r%
\rightarrow0,\;\forall F\in D_{h^{\ast}}, \\
& u_{\pm}=\left( 
\begin{array}{c}
1 \\ 
\frac{\varkappa\pm i\sigma}{q}%
\end{array}
\right) ,
\end{align*}
which yields 
\begin{equation*}
\Delta_{\ast}(F)=\frac{2i\sigma}{q}(|c_{1}|^{2}-|c_{2}|^{2}).
\end{equation*}

It follows that in the case of overcritical charges, i.e., for $\Upsilon
_{+}=i\sigma $, we have the one-parameter $U(1)$-family $\{\hat{h}_{\theta
}\,\}$, $0\leq \theta \leq \pi $, $0\sim \pi $, of s.a. operators associated
with s.a. differential expression $\check{h}$ (\ref{R.4}) and specified by
s.a. asymptotic boundary conditions \footnote{%
The relation $c_{2}=e^{i\theta }c_{1},\;0\leq \theta \leq 2\pi ,$ defining
s.a. boundary conditions (compare with (\ref{3.3.1.2})), is equivalent to
the relations $c_{1}=e^{i\theta }c,\;c_{2}=e^{-i\theta }c,\;0\leq \theta
\leq \pi ,$ with the change $\theta \rightarrow 2\pi -2\theta .$} 
\begin{equation}
\hat{h}_{\theta }:\left\{ 
\begin{array}{l}
D_{\theta }=\left\{ 
\begin{array}{l}
F(r):F(r)\;\text{\textrm{is absolutely continuous in} }(0,\infty ),\;F,%
\check{h}F\subset \mathcal{L}^{2}(0,\infty ), \\ 
F(r)=c[e^{i\theta }(mr)^{i\sigma }u_{+}+e^{-i\theta }(mr)^{-i\sigma
}u_{-}+O(r^{1/2}),\;r\rightarrow 0, \\ 
0\leq \theta \leq \pi ,\;0\sim \pi ,%
\end{array}%
\right\} \\ 
\,\hat{h}_{\theta }F=\check{h}F,%
\end{array}%
\right.  \label{3.5.1.1}
\end{equation}%
$D_{\theta }$ is the domain of $\hat{h}_{\theta }$.

\subsubsection{Spectral analysis}

\label{sec4.5.2}

For the doublet $U(r;W)$\ defining guiding functional (\ref{A1.0}), we
choose the doublet 
\begin{equation*}
U_{\theta }(r;W)=e^{i\theta }U_{(1)}(r;W)+e^{-i\theta }U_{(2)}(r;W)
\end{equation*}%
with the asymptotic behavior%
\begin{equation*}
U(r;W)=e^{i\theta }(mr)^{i\sigma }u_{+}+e^{-i\theta }(mr)^{-i\sigma
}u_{-}+O(r),\;r\rightarrow 0,
\end{equation*}%
where $U_{(1)\text{ }}$and $U_{(2)}$ are given by formulas (\ref{1.27}) -- (%
\ref{1.33}) (with $\Upsilon _{+}=i\sigma $ ); $U_{\theta }(r;W)$ is
real-entire in $W$ because $U_{(2)}=\overline{U_{(1)}}$ for $\Upsilon
_{+}=i\sigma $ and real $W=E.$

For $\mathcal{D}$, we choose the set $\mathcal{D}_{\theta }$ of the doublets
belonging to $D_{\theta }$ and having a compact support.

Using the doublets $U_{\theta }(r;E_{0})$ and $U_{(1)}(r;E_{0})$ for
constructing the solution $\Psi\in\mathcal{D}_{\theta }$ of the
inhomogeneous equation $(\check{h}-E_{0})\Psi =F_{0}$, $F_{0}\in \mathcal{D}%
_{\theta }$, we verify that the guiding functional $\Phi _{\theta }$ is
simple.

Green's function $G_{\theta }(r,r^{\prime };W)$, $\Im W>0$, of the
Hamiltonian $\hat{h}_{\theta }$ is constructed in terms of the doublets $%
U=U_{\theta }$ and $V=V_{\theta }$,\ where 
\begin{align*}
& V_{\theta }(r;W)=\frac{2}{e^{-i\theta }+e^{i\theta }\tilde{\omega}(W)}%
V_{(1)}(r;W)=U_{\theta }(r;W)-\frac{q}{4\sigma }\omega _{\theta }(W)\tilde{U}%
_{\theta }(r;W), \\
& \tilde{U}_{\theta }(r;W)=\frac{1}{i}[e^{i\theta }U_{(1)}(r;W)-e^{-i\theta
}U_{(2)}(r;W)], \\
& \tilde{\omega}(W)=\frac{q}{2i\sigma }\omega (W),\;\omega _{\theta }(W)=-%
\mathrm{Wr}(U,V)=-\frac{4i\sigma }{q}\frac{1-\tilde{\omega}(W)e^{2i\theta }}{%
1+\tilde{\omega}(W)e^{2i\theta }},
\end{align*}%
$V_{(1)}(r;W)$ and $\omega (W)$ are given in(\ref{1.34}) -- (\ref{1.36})
(with $\Upsilon _{+}=i\sigma $); $\tilde{U}_{\theta }(r;W)$ is real-entire
in $W$. As a result, we obtain that 
\begin{equation*}
G_{\theta }(r,r^{\prime };W)=\left\{ 
\begin{array}{c}
\frac{1}{\omega _{\theta }(W)}V_{\theta }(r;W)\otimes U_{\theta }(r^{\prime
};W),\;r>r^{\prime }, \\ 
\frac{1}{\omega _{\theta }(W)}U_{\theta }(r;W)\otimes V_{\theta }(r^{\prime
};W),\;r<r^{\prime }.%
\end{array}%
\right.
\end{equation*}%
The spectral function $\sigma _{\theta }(E)$ of the radial Hamiltonian $%
\hat{h}_{\theta }$ is defined by 
\begin{equation*}
\frac{d\sigma _{\theta }(E)}{dE}=\frac{1}{\pi }\lim_{\varepsilon \rightarrow
0}\Im\frac{1}{\omega _{\theta }(E+i\varepsilon )}
\end{equation*}%
and is determined by the (generalized) function $\Im\omega^{-1}(E)$, 
\begin{equation*}
\omega _{\theta }^{-1}(E)=\lim_{\varepsilon \rightarrow 0}\frac{1}{\omega
_{\theta }(E+i\varepsilon )}.
\end{equation*}%
At the points where the function 
\begin{equation*}
\omega _{\theta }(E)=\lim_{\varepsilon \rightarrow 0}\omega _{\theta
}(E+i\varepsilon )
\end{equation*}%
is different from zero, we have $\omega _{\theta }^{-1}(E)=1/\omega _{\theta
}(E)$.

The two energy regions $|E|\geq m\,$and $|E|<m\,$are naturally distinguished
as before.

In the region $|E|\geq m$, the function $\omega _{\theta }(E)$ is
continuous, different from zero, and complex, therefore, the spectral
function $\sigma _{\theta }(E)$ for $|E|\geq m$ is absolutely continuous,
and 
\begin{equation*}
\frac{d\sigma _{\theta }(E)}{dE}=\frac{1}{\pi }\Im\frac{1}{\omega _{\theta
}(E)}\equiv Q_{\theta }^{2}(E).
\end{equation*}

In the region $\ |E|<m$, we have 
\begin{equation*}
\tilde{\omega}(E)=\frac{\Gamma (2i\sigma )}{\Gamma (-2i\sigma )}\frac{\Gamma
(-i\sigma -Eq/\tau )}{\Gamma (i\sigma -Eq/\tau )}\frac{\tau (\varkappa
+i\sigma )-q(m-E)}{\tau (\varkappa -i\sigma )-q(m-E)}(2\tau /m)^{-2i\sigma
}\equiv e^{-2i\Theta (E)},
\end{equation*}%
therefore the function 
\begin{equation*}
\omega _{\theta }(E)=\frac{4\sigma }{q}\tan (\Theta (E)-\theta )
\end{equation*}%
is real.

It follows that the spectral function $\sigma _{\theta }(E)$ for $\ |E|<m$
is a jump function with the jumps $Q_{\theta ,n}^{2}$, 
\begin{equation*}
Q_{\theta ,n}^{2}=\lim_{E\rightarrow E_{\theta ,n}}\frac{E_{\theta ,n}-E}{%
\omega _{\theta }(E)},
\end{equation*}%
located at the discrete points $E_{\theta ,n}$\ where $\omega _{\theta
}(E_{\theta ,n})=0$, such that 
\begin{equation*}
\frac{d\sigma _{\theta }(E)}{dE}=\sum_{n}Q_{\theta ,n}^{2}\delta
(E-E_{\theta ,n}).
\end{equation*}%
We failed to find an explicit formula for the discrete energy eigenvalues $%
E_{\theta ,n}.$ We only note that there is an infinite number of the
discrete levels which accumulate at the point $E=m$. Their asymptotic form
as $n\rightarrow \infty $ is given by the previous nonrelativistic formula: 
\begin{equation*}
\epsilon _{\theta ,n}\equiv m-E_{\theta ,n}=\frac{mq^{2}}{2n^{2}}.
\end{equation*}
The lower bound state energy essentially depends on $\theta $, and there
exists a value of $\theta $ for which the lower bound state energy coincides
with the boundary $E=-m$ of the lower (positron) continuous spectrum.

The inversion formulas and Parseval equality in terms of normalized
(generalized) eigenfunctions $U_{\theta ,\mathrm{norm}}(r;E)$\ and the
Fourier coefficients $\varphi _{\theta }(E)\;$are copies of formulas (\ref%
{3.1.4.3c}), (\ref{3.1.4.3d}), (\ref{3.1.4.4}), (\ref{3.1.4.5}), and (\ref%
{3.1.4.5a}).

As in the previous subsections, a comment on the applicability of the
standard method for finding the energy spectrum and eigenfunctions based on
the physical arguments holds. As an example, we consider the energy region $%
|E|<m$ where the solutions $F$ of the differential equation $\check{h}F=EF$
either exponentially grow or exponentially decrease as $r\rightarrow \infty $%
, any solution being square integrable at the origin. Only exponentially
decreasing solutions 
\begin{equation*}
F=cV_{\theta }(r;W)=c[U_{\theta }(r;W)-\frac{q}{4\sigma }\omega _{\theta }(W)%
\tilde{U}_{\theta }(r;W)],
\end{equation*}%
$c$ is a constant, are proper. They are square integrable, $F\in \mathcal{L}%
^{2}(0,\infty )$, for any $E\;$\ in the interval $|E|<m$, but satisfy s.a.
asymptotic boundary conditions (\ref{3.5.1.1})\ only if $\omega _{\theta
}(E)=0$, which reproduces the results for the eigenvalues and eigenfunctions
of the discrete spectrum. We can also establish the orthonormality relations
for the eigenfunctions $U_{\theta ,\mathrm{norm}}(r;E)$, which are analogues
of relations (\ref{3.1.4.5b}), by a direct calculation of the corresponding
integrals with the method described in Appendix~\ref{appB}.

In conclusion, we point out that the number of s.a. extensions of the total
Dirac Hamiltonian (i.e., the number of independent parameters of s.a.
extensions) depends on the charge values $q$ (on the value of $Z$) as
follows: it is easy to verify that in the interval $q_{n}<q\leq q_{n+1}$,
where 
\begin{equation*}
q_{n}=\left\{ 
\begin{array}{l}
0,\quad n=0, \\ 
\sqrt{n^{2}-1/4},\quad n=1,2,...,%
\end{array}%
\right.
\end{equation*}%
the number of independent parameters of s.a. extensions is equal to $2n$.
This follows from the fact that the total Dirac Hamiltonian is a direct sum
of its parts unitary equivalent to the radial Hamiltonians, see (\ref{R.7}),
(\ref{R.6}).

\section*{Appendix A}

\label{appA}

\setcounter{equation}{0}

One of the methods for finding the spectrum of a s.a. differential operator
and its complete system of (generalized) eigenfunctions and constructing the
corresponding Fourier expansion with respect to these eigenfunctions (the
so-called inversion formulas) is based on the Krein method of guiding
functionals. For ordinary scalar differential operators, this method is
described in~\cite{18}, but it is directly extended to the case of ordinary
matrix operators. We here present the key points of the method as applied to
our case where it suffices to consider only one guiding functional. This
implies that the operator spectrum is simple\footnote{%
For a definition, see~\cite{18}. In the physical terminology, this means
that the spectrum is nondegenerate.}; we call such a functional a simple
guiding functional.

By definition, a guiding functional $\Phi (F;W)$ for a s. a. operator $%
\hat{h}$\ \ associated with a differential expression $\check{h}$ is a
functional of the form 
\begin{equation}
\Phi (F;W)=\int_{0}^{\infty }U(r;W)F(r)dr,  \label{A1.0}
\end{equation}%
where$\;U(r;W)$ is a solution of the homogeneous equation 
\begin{equation*}
(\check{h}-W)U=0
\end{equation*}%
which is real-entire\ in $W$ and $F(r)$ belongs to some subspace $\mathcal{D}%
\subset D\cap D_{h}$, where $D$ is a space of doublets with a compact
support such that $\mathcal{D}$ is dense in $\mathcal{L}^{2}(0,\infty )$.

A guiding functional $\Phi (F;W)\;$is called simple if it satisfies the
following conditions:

1) for a fixed $F$, the functional $\Phi(F;W)$ is an entire function of $W$;

2) if 
\begin{equation*}
\Phi(F_{0};E_{0})=0,\;\Im E_{0}=0,\;F_{0}\in\mathcal{D},
\end{equation*}
then the inhomogeneous\ equation 
\begin{equation*}
(\check{h}-E_{0})\Psi=F_{0}
\end{equation*}
has a solution $\Psi\in\mathcal{D}$;

3) $\Phi(\hat{h}F;W)=W\Phi(F;W)$.

We note, that the existence of a simple guiding functional is conditioned by
the existence of appropriate $U$\ and $\mathcal{D}$. A guiding principle for
the choice of $U(r;W)$ is that its behavior as $r\rightarrow 0$ must conform
to the asymptotic behavior admissible for the doublets belonging to $D_{h}$,
or, roughly speaking, $U(r;W)$ at the origin \ must belong to $D_{h}$. This
corresponds to the conventional physical requirement that the (generalized)
eigenfunctions of the operator $\hat{h}$, being generally
non-square-integrable, but ``normalizable to $\delta $-function'', satisfy
the s.a. boundary conditions specifying $\hat{h}$.

If a simple guiding functional exists, then the s. a. operator $\hat{h}$ has
the following spectral properties:

1) the spectrum of $\hat{h}$ is simple, and there exists a spectral function 
$\sigma (E)$, a nondecreasing real function continuous from the right and
such that the set of spectrum points coincides with the set of growth points%
\footnote{%
The set of growth points of the function $\sigma (E)$ is the complement of
the open set of constancy points of the function $\sigma (E)$. A point $%
E_{0} $ is the constancy point of the function $\sigma (E)$ if there exists
a vicinity of the point $E_{0}$ where $\sigma (E)$ is constant.} of the
function $\sigma (E)$;

2) the inversion formulas 
\begin{align}
& \Phi (E)=\int_{0}^{\infty }U(r;E)F(r)dr,  \label{A1.1} \\
& F(r)=\int_{-\infty }^{\infty }U(r;E)\Phi (E)d\sigma (E),  \label{A1.2}
\end{align}%
and the Parseval\emph{\ }equality 
\begin{equation}
\int_{0}^{\infty }|F(r)|^{2}dr=\int_{-\infty }^{\infty }|\Phi
(E)|^{2}d\sigma (E)  \label{A1.2ab}
\end{equation}%
hold, where $\Phi (E)\in L_{\sigma }^{2}(-\infty ,\infty )$, $F(r)\in 
\mathcal{L}^{2}(0,\infty )$, the function $U(r;E)\;$in the integrands in (%
\ref{A1.1}) and (\ref{A1.2}) can be defined as zero outside of the spectrum
points of the operator $\hat{h}$ (outside of the growth points of $\sigma
(E) $), and the convergence of the integrals in (\ref{A1.1}) and (\ref{A1.2}%
) in general must be understood in the sense of the convergence with respect
to the metrics of the respective spaces $L_{\sigma }^{2}(-\infty ,\infty)$
and $\mathcal{L}^{2}(0,\infty)$. This means that the set of the
(generalized) eigenfunctions $\{U(r;E),E\in\mathrm{Spec}\hat{h}\}$ of the
operator $\hat{h} $ forms a complete orthogonal system.

The spectral function $\sigma (E)\;$can \ be expressed via the resolvent of
the operator $\hat{h}$. As is known, see~\cite{18}, the resolvent $\hat{R}%
(W)=(\hat{h}-W)^{-1}\;$with $\Im W\neq 0$ is an integral operator with the
kernel $G(r,r^{\prime };W)$ ( Green's function). The spectral function $%
\sigma (E)$ is expressed in terms of Green's function as follows: 
\begin{align}
& U(c;E)\otimes U(c;E)d\sigma (E)=d\mathcal{M}(c;E),  \label{A1.3a} \\
& \mathcal{M}(c;E)=\lim_{\delta \rightarrow +0}\lim_{\varepsilon \rightarrow
+0}\frac{1}{\pi }\int_{\delta }^{E+\delta }\Im M(c;E^{\prime }+i\varepsilon
)dE^{\prime },  \label{A1.3b} \\
& M(c;W)=G(c-0,c+0;W),  \label{A1.3c}
\end{align}%
where $c\;$is an arbitrary internal point of the interval $(0,\infty )$. We
note that for any $E$, one of the diagonal elements of the matrix $%
U(c;E)\otimes U(c;E)$ is different from zero. Of course, $\sigma (E)$ is
independent of $c$.

\section*{Appendix B}

\label{appB}

\setcounter{equation}{0}

In this Appendix, we describe a method for calculating the so-called overlap
integrals for the solutions of the differential equation $\check{h}F=EF$ .
It is based on the integral Lagrange identity and on evaluating the
asymptotic behavior of the solutions at the boundaries, the origin and
infinity. We apply this method to proving the orthonormality relations for
the (generalized) eigenfunctions of the radial Hamiltonian and illustrate it
by the example of the second noncritical charge region.

We call the integral\footnote{%
In this Appendix, a notation like $F^{\prime }$ denotes another function,
but not a derivative.} 
\begin{equation*}
\int_{0}^{\infty }F(r;W)F^{\prime }(r;W^{\prime })dr=\lim_{R\rightarrow
\infty ,\epsilon \rightarrow 0}\int_{\epsilon }^{R}F(r;W)F^{\prime
}(r;W^{\prime })dr
\end{equation*}%
for two doublets $F\,$and $F$ $^{\prime }$ the overlap integral for these
doublets. Let $F\,$and $F$ $^{\prime }$ satisfy the respective homogeneous
equations 
\begin{equation*}
(\check{h}-W)F(r;W)=0\;\mathrm{and}\;(\check{h}-W^{\prime })F^{\prime
}(r;W^{\prime })=0.
\end{equation*}%
Then the equality for the overlap integral 
\begin{equation}
\int_{0}^{\infty }F(r;W)F^{\prime }(r;W^{\prime })dr=I^{\infty }-I_{0},
\label{A2.3.1}
\end{equation}%
where%
\begin{eqnarray}
I^{\infty } &=&\lim_{r\rightarrow \infty }\frac{\mathrm{Wr}(r;F,F^{\prime })%
}{W-W^{\prime }},  \label{A2.3.2} \\
I_{0} &=&\lim_{r\rightarrow 0}\frac{\mathrm{Wr}(r;F,F^{\prime })}{%
W-W^{\prime }}  \label{A2.3.3}
\end{eqnarray}%
holds. The equality (\ref{A2.3.1}) is a special case of the integral
Lagrange identity. In what follows, we are interested in the case of real $W$
and $W^{\prime }$, $W=E$ and $W^{\prime }=E^{\prime }$. The overlap integral
is understood as a generalized function of $E$ and $E^{\prime }$. Evaluating
the overlap integrals is thus reduced to evaluating the asymptotics of the
Wronskian of the corresponding doublets at the boundaries.

We begin with evaluating the asymptotics of some basic functions and
doublets.

Let $|E|\geq m$, in which case we have $K=|E|k/E$, $k=\sqrt{E^{2}-m^{2}}$,
and let $r\rightarrow\infty$. Using the known asymptotics of the functions $%
\Phi(\alpha,\beta;x)$ (see, for example,~\cite{21}), we have 
\begin{align*}
& \Phi(\alpha,\beta;2i\xi_{1}kr)\rightarrow\frac{\Gamma(1+2\Upsilon)}{%
\Gamma(1+\Upsilon-i\xi_{2}qE/k)}e^{i\xi_{1}\pi\Upsilon/2}e^{-\xi_{1}\xi
_{2}\pi Eq/(2k)}(2kr)^{-\Upsilon-i\xi_{2}Eq/k}, \\
& \alpha=\Upsilon+i\xi_{2}qE/k,\;\xi_{1}=\pm1,\;\xi_{2}=\pm1,\;\beta
=1+2\Upsilon,
\end{align*}
where $\Upsilon$ is any real or pure imaginary number, $\Upsilon\neq-n/2$, $%
n=1,2,...$. If $\Upsilon$ is real, $\Upsilon=\gamma\neq-n/2$, $n=1,2,...$,
then 
\begin{align*}
& (mr)^{\Upsilon}\Phi_{+}(r,\Upsilon,E,k)\rightarrow2\Delta(\Upsilon
,E)\cos\psi(r;\Upsilon,E), \\
& (mr)^{\Upsilon}\Phi_{-}(r,\Upsilon,E,k)\rightarrow\frac{2}{k}%
\Delta(\Upsilon,E)\sin\psi(r;\Upsilon,E),
\end{align*}
where 
\begin{align*}
&\Delta(\Upsilon,E)=\frac{\Gamma(1+2\Upsilon)(2k/m)^{-\Upsilon} e^{-\pi
qE/(2k)}}{|\Gamma(1+\Upsilon+iqE/k)|}, \\
&\psi(r;\Upsilon,E)=kr+\frac{qE}{k}\ln(2kr)-\frac{\pi\Upsilon}{2}
-\psi_\Gamma(\Upsilon,E), \\
&\psi_\Gamma(\Upsilon,E)=\arg{\Gamma(1+\Upsilon+iqE/k)},
\end{align*}
which yields 
\begin{equation*}
X(r,\Upsilon,E,k)\rightarrow\Delta(\Upsilon,E)\left[ \cos\psi(r;%
\Upsilon,E)u_{+}+\frac{\sin\psi(r;\Upsilon,E)}{k}\left( 
\begin{matrix}
\frac{(m+W)(\varkappa+\gamma)}{q} \\ 
m-E%
\end{matrix}
\right)\right].
\end{equation*}

Let $|E|<m$. For our purposes, it suffices to know that the doublets $%
U(r;E_{n})$ and $V(r;E)$ decrease exponentially as $r\rightarrow\infty$.

We use the relation 
\begin{align}
& \frac{E}{|E|}\frac{\sin[\psi_{\Gamma}(r;\Upsilon,E)\pm\psi_{\Gamma
}(r;\Upsilon^{\prime},E^{\prime})}{E-E^{\prime}}\rightarrow\frac{|E|}{k}%
\frac{\sin[(k-k^{\prime})r}{k-k^{\prime}}\rightarrow  \notag \\
& \,\rightarrow\frac{|E|}{k}\pi\delta(k-k^{\prime})=\pi\delta(E-E^{\prime
}),\;r\rightarrow\infty,  \label{A2.3.4}
\end{align}
which holds in a distribution-theoretic sense. We call the expressions of
the form 
\begin{align*}
&a_{\pm}(E,E^{\prime})\sin[\psi_{\Gamma}(r;\Upsilon,E)\pm
\psi_{\Gamma}(r;\Upsilon^{\prime},E^{\prime})], \\
& b_{\pm}(E,E^{\prime})\cos[\psi_{\Gamma}(r;\Upsilon,E)\pm\psi_{\Gamma
}(r;\Upsilon^{\prime},E^{\prime})],
\end{align*}
where $a_{\pm}(E,E^{\prime})$, $b_{\pm}(E,E^{\prime})$ are finite at $%
E=E^{\prime}$, the quickly oscillating expressions (QO); such expressions
have the zero limit in a distribution-theoretic sense as $r\rightarrow\infty$%
.

This allows obtaining the limits $I^{\infty}$ for the basic doublets. We
have 
\begin{align}
&
U_{(1)}=U_{(1)}(r;\gamma,E),\;U_{(1)}^{\prime}=U_{(1)}^{\prime}(r;\gamma,E^{%
\prime}),\;\gamma>0,  \notag \\
& \,(E-E^{\prime})^{-1}W(r;U_{(1)},U_{(1)}^{\prime})\rightarrow  \notag \\
& \,\rightarrow A(\gamma,E)\frac{E}{|E|}\frac{\sin[\psi_{\Gamma}(\gamma,E)-%
\psi_{\Gamma}(\gamma,E^{\prime})]}{\pi(E-E^{\prime})}+\mathrm{QO}\rightarrow
\notag \\
& \,\rightarrow A(\gamma,E)\delta(E-E^{\prime}),\;r\rightarrow\infty
\label{A2.3.1.1}
\end{align}%
\begin{align}
&
U_{(1)}=U_{(1)}(r;\gamma,E),\;U_{(2)}^{\prime}=U_{(2)}^{\prime}(r;\gamma,E^{%
\prime}),\;0<\gamma<1/2,  \notag \\
& \,(E-E^{\prime})^{-1}W(r;U_{(1)},U_{(2)}^{\prime})\rightarrow  \notag \\
& \,\rightarrow B(\gamma,E)\frac{E}{|E|}\frac{\sin[\psi_{\Gamma}(\gamma,E)-%
\psi_{\Gamma}(-\gamma,E^{\prime})]}{\pi(E-E^{\prime})}-  \notag \\
& \,-B(\gamma,E)\frac{\gamma k}{q|E|}\frac{\cos[\psi_{\Gamma}(\gamma
,E)-\psi_{\Gamma}(-\gamma,E^{\prime})]}{\pi(E-E^{\prime})}+\mathrm{QO}%
\rightarrow  \notag \\
& \,\rightarrow B(\gamma,E)\delta(E-E^{\prime})-B(\gamma,E)\frac{\gamma k}{%
q|E|}\frac{\cos[\psi_{\Gamma}(\gamma,E)-\psi_{\Gamma}(-\gamma,E^{\prime})]}{%
\pi(E-E^{\prime})},\;r\rightarrow\infty,  \label{A2.3.1.2}
\end{align}%
\begin{align}
&
U_{(2)}=U_{(2)}(r;\gamma,E),\;U_{(1)}^{\prime}=U_{(1)}^{\prime}(r;\gamma,E^{%
\prime}),\;0<\gamma<1/2,  \notag \\
& \,(E-E^{\prime})^{-1}W(r;U_{(2)},U_{(1)}^{\prime})\rightarrow  \notag \\
& \,\rightarrow B(\gamma,E)\delta(E^{\prime}-E)-B(\gamma,E)\frac{\gamma k}{%
q|E|}\frac{\cos[\psi_{\Gamma}(\gamma,E^{\prime})-\psi_{\Gamma}(-\gamma ,E)]}{%
\pi(E^{\prime}-E)}\rightarrow  \notag \\
& \,\rightarrow B(\gamma,E)\delta(E-E^{\prime})+B(\gamma,E)\frac{\gamma k}{%
q|E|}\frac{\cos[\psi_{\Gamma}(\gamma,E)-\psi_{\Gamma}(-\gamma,E^{\prime})]}{%
\pi(E-E^{\prime})},\;r\rightarrow\infty,  \label{A2.3.1.3}
\end{align}%
\begin{align}
&
U_{(2)}=U_{(2)}(r;\gamma,E),\;U_{(2)}^{\prime}=U_{(2)}^{\prime}(r;\gamma,E^{%
\prime}),\;0<\gamma<1/2,  \notag \\
& \,(E-E^{\prime})^{-1}W(r;U_{(2)},U_{(2)}^{\prime})\rightarrow  \notag \\
& \,\rightarrow A(-\gamma,E)\frac{E}{|E|}\frac{\sin[\psi_{\Gamma}(-%
\gamma,E)-\psi_{\Gamma}(-\gamma,E^{\prime})]}{\pi(E-E^{\prime})}+\mathrm{QO}%
\rightarrow  \notag \\
& \,\rightarrow A(-\gamma,E)\delta(E-E^{\prime}),\;r\rightarrow \infty,
\label{A2.3.1.4}
\end{align}%
where 
\begin{align}
A(\Upsilon,E) & =\Delta^{2}(\Upsilon,E)\frac{2\pi(q_{cj}+\zeta\Upsilon
)(q_{cj}|E|+\zeta m\Upsilon)}{kq^{2}},  \label{A2.3.1.5} \\
B(\gamma,E) & =\frac{|E|}{k}\Delta(\gamma,E)\Delta(-\gamma ,E).
\label{A2.3.1.6}
\end{align}
For the corresponding limits $I_{0}$, we respectively find (the relations
are valid for any $E$, $E^{\prime}$) 
\begin{align}
&U_{(1)}=U_{(1)}(r;\gamma,E),\;U_{(1)}^{\prime}=
U_{(1)}^{\prime}(r;\gamma,E^{\prime}),\;\gamma>0,  \notag \\
& \,(E-E^{\prime})^{-1}W(r;U_{(1)},U_{(1)}^{\prime})\rightarrow
0,\;r\rightarrow0,  \label{A2.3.2.1}
\end{align}%
\begin{align}
&U_{(1)}=U_{(1)}(r;\gamma,E),\;U_{(2)}^{\prime}=
U_{(2)}^{\prime}(r;\gamma,E^{\prime}),\;0<\gamma<1/2,  \notag \\
&(E-E^{\prime})^{-1}W(r;U_{(1)},U_{(2)}^{\prime})\rightarrow- \frac{2\gamma}{%
q(E-E^{\prime})},\;r\rightarrow0,  \label{A2.3.2.2}
\end{align}%
\begin{align}
&
U_{(2)}=U_{(2)}(r;\gamma,E),\;U_{(1)}^{\prime}=U_{(1)}^{\prime}(r;\gamma,E^{%
\prime}),\;0<\gamma<1/2,  \notag \\
&(E-E^{\prime})^{-1}W(r;U_{(2)},U_{(1)}^{\prime})\rightarrow\frac{2\gamma }{%
q(E-E^{\prime})},\;r\rightarrow0,  \label{A2.3.2.3}
\end{align}%
\begin{align}
&U_{(2)}=U_{(2)}(r;\gamma,E),\;U_{(2)}^{\prime}=
U_{(2)}^{\prime}(r;\gamma,E^{\prime}),\;0<\gamma<1/2,  \notag \\
& \,(E-E^{\prime})^{-1}W(r;U_{(2)},U_{(2)}^{\prime})\rightarrow
0,\;r\rightarrow0.  \label{A2.3.2.4}
\end{align}

The obtained relations allow calculating the overlap integrals and proving
the orthonormality relations for the eigenfunctions of the radial
Hamiltonians. As an example, we consider the second noncritical charge
region, the other charge regions, including the critical and overcritical
regions, are considered quite similarly.

We have to calculate the integral 
\begin{equation*}
\int_{0}^{\infty}U_{\xi}(r;E)U_{\xi}(r;E^{\prime})dr,
\end{equation*}
where $U_{\xi}(r;E)$ is defined by Eq. (\ref{3.3.2.1}). Using relations (\ref%
{A2.3.1}) -- (\ref{A2.3.3}) and (\ref{A2.3.1.1}) -- (\ref{A2.3.2.4}), we
find 
\begin{align*}
& \int_{0}^{\infty}U_{\xi}(r;E)U_{\xi}(r;E^{\prime})dr=C_{\xi}(E)\delta
(E-E^{\prime}),\;|E|,|E^{\prime}|\geq m, \\
& C_{\xi}(E)=A(\gamma,E)+2\xi B(\gamma,E)+\xi^{2}A(-\gamma,E),
\end{align*}%
\begin{equation*}
\int_{0}^{\infty}U_{\xi}(r;E_{\xi,n})U_{\xi}(r;E^{\prime})dr=0,\;|E^{\prime
}|\geq m,
\end{equation*}%
\begin{equation*}
\int_{0}^{\infty}U_{\xi}(r;E_{\xi,n})U_{\xi}(r;E_{\xi,n^{\prime}})dr=0,\;n%
\neq n^{\prime}.
\end{equation*}

The normalization factor $A_{n}$ for the eigenfunctions of the discrete
spectrum, 
\begin{equation*}
A_{n}^{2}=\int_{0}^{\infty}U_{\xi}^{2}(r;E_{\xi,n})dr,
\end{equation*}
can also be calculated (see, for example,~\cite{14}). It is interesting to
note that we can explicitly verify the fulfillment of the relation $%
A_{n}=Q_{\xi,n}^{-1}$. For this purpose, we consider the integral 
\begin{equation*}
\int_{0}^{\infty}U_{\xi}(r;E_{\xi,n})V(r;E^{\prime})dr=\lim_{r\rightarrow 0}%
\frac{\mathrm{Wr}(r;U_{\xi},V^{\prime})}{E^{\prime}-E_{\xi,n}},\;V^{\prime
}=V(r;E^{\prime}),\;|E^{\prime}|<m,
\end{equation*}
where $V(r;E)$ is defined by (\ref{3.3.2.2}). Using relations (\ref{A2.3.2.1}%
) -- (\ref{A2.3.2.4}), we find 
\begin{equation*}
\int_{0}^{\infty}U_{\xi}(r;E_{\xi,n})V(r;E^{\prime})dr=\frac{\omega_{\xi
}(E^{\prime})}{E_{\xi,n}-E^{\prime}}.
\end{equation*}
We now recall that $V(r;E_{\xi,n})=U_{\xi}(r;E_{\xi,n})$ and finally obtain
that 
\begin{equation*}
\int_{0}^{\infty}U_{\xi}^{2}(r;E_{\xi,n})dr=\lim_{E^{\prime}\rightarrow
E_{\xi,n}}\frac{\omega_{\xi}(E^{\prime})}{E_{\xi,n}-E^{\prime}}%
=Q_{\xi,n}^{-2}.
\end{equation*}

We also note that it follows from the inversion formulas that $C_{\xi
}(E)=Q_{\xi}^{-2}(E)$.

\subsection*{Acknowledgement}

Gitman is grateful to the Brazilian foundations FAPESP and CNPq for
permanent support; Voronov thanks FAPESP for support during his stay in
Brazil; Tyutin and Voronov thank RFBR, grants 05-01-00996(IT) and
05-02-17471(BV), and LSS-4401.2006.2 for partial support.

\end{document}